# Foundations of Nonlinear Dynamics
# or
# Real Dynamics, Ideal Dynamics, Unpredictable Dynamics and the "Schroedinger's Cat Paradox"

Kupervasser Oleg
Technion, Haifa, Israel

**Abstract**

*The paper discusses the basic paradoxes of thermodynamics and quantum mechanics. The approaches to solution of these paradoxes are suggested. The first one relies on the influence of the external observer (environment), which disrupts the correlations in the system. The second one is based on the limits of self-knowledge of the system in case of both the external observer and the environment is included in the considered system. The concepts of Real Dynamics, Ideal Dynamics, and Unpredictable dynamics are introduced. The phenomenon of Life is contemplated from the point of view of these Dynamics.*

## Introduction.

Statistical mechanics and quantum mechanics are the well-developed and well-known theories lying at the base of modern physics. However, they contain a number of paradoxes that cause many physicists to doubt the consistency of these theories. Below a brief overview of these paradoxes is given.

1) Basic equations of classical and quantum mechanics are time-reversible. However, the law of entropy increase in closed systems, which should be deducible from these equations, is clearly non-reversible in time [1, 2].
2) The same relates to the contradiction between the law of entropy increase in closed systems (see Appendix E) and the Poincare's theorem. The latter states that a closed system can return to a point, which is arbitrarily close to the initial state of the system [1, 2].
3) The Schroedinger's "Cat paradox", i.e. the reduction of wave packet (see Appendix B) and the transition between pure and mixed states in macroscopic systems or during the measurement of the quantum system. This process cannot be described by the quantum mechanical equations. There is absolutely no clarity as to the time moment when the process occurs and what is its duration. However, quantum mechanics is incomplete without this process [3].
4) Microscopic entropy (see Appendix E) of classical systems determined via the state density function in phase space remains constant. The same is true for the entropy of quantum mechanical (QM) systems determined via the density matrix [15]. Macroscopic entropy (see Appendix E) of the systems determined in terms of macroscopic variables or coarsened (see Appendix D) phase density function (see Appendix C) in the case of Classical mechanics (or the reduced density matrix (see Appendix A) in case of QM) can both increase and decrease. Both the processes have equal probabilities due to the reversibility of motion. This contradicts the sometimes expressed mistaken view that the processes accompanied by the entropy increase to be more probable than those accompanied by its decrease. Why then the increase of entropy is only observed in all the real macroscopic systems? If the basic equations of classical mechanics should provide the complete description of the system, why then the introduction of additional assumptions is necessary to derive the law of entropy increase? [1, 2] (The Boltzmann's molecular chaos hypothesis and the equivalent roughening (see Appendix D) of the density function in phase state (see Appendix C) are among such additional assumptions.)

5) Various sets of macroscopic variables can be used for describing the macroscopic state of the system. Theoretically, there are no limitations on the choice of the macroscopic variables; they can be arbitrary functions of the microscopic parameters. However, the number of the "convenient" variables is limited. Why?
6) There are a great number of live organisms, including those that are self-aware and possess freedom of will. Which physical properties constitute the principal difference between live (self-aware) systems from non-live ones? Can the former be completely described by the known physical equations?

Note that Schroedinger used a live cat as a macroscopic detector in formulating his paradox. A certain correlation between the basic physical paradoxes and the enigma of life is intuitively obvious for many physicists. We shall try to formulate this correlation more clearly.

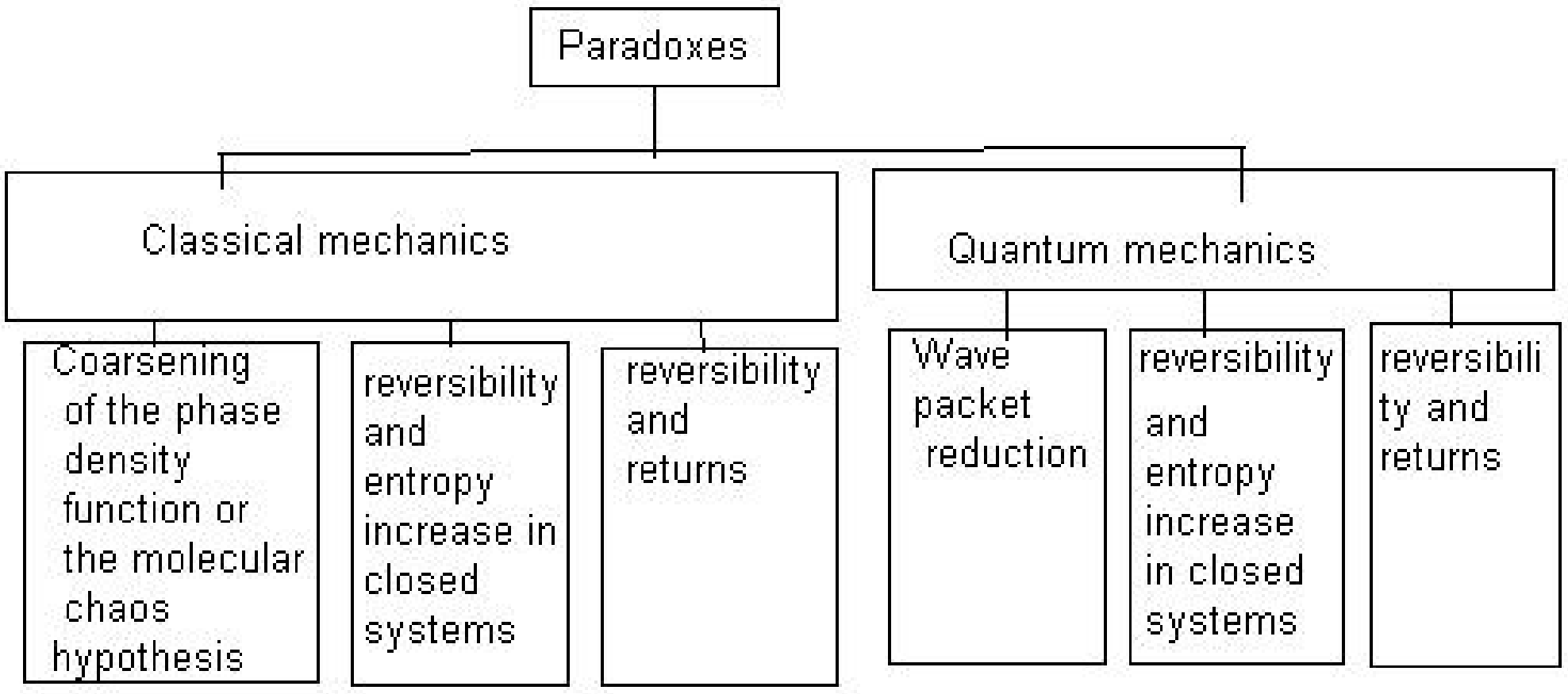

Figure1 Paradoxes in Classical and Quantum mechanics

The solution of the abovementioned paradoxes does not require introduction of new physical principles. However, the experimental testing of these paradoxes can lead to discovering new laws of physics.

We shall analyze these problems remaining strictly within the limits of the well-known physical laws.

A number of publications treat on these paradoxes (See, e.g. [1, 2, 3]). Most of the conclusions are quite true. The problem with most of the approaches used previously is that they do not analyze the key property of the dynamic system that leads to these paradoxes. Thus they are able to give only partial solutions of the paradoxes.

The situation with these approaches reminds the attempts to solve the Gibbs' paradox [4]. The step-like entropy increase on mixing of different gases was often explained by the absence of intermediate forms between identical and different gases. However, such intermediate forms are feasible, and thus the value of the step can be controlled. Thus a more thorough consideration of this problem yields both a more deep understanding of this paradox, and a deeper insight into the fundamentals of physics, while requiring introducing no new physical laws.

The same is true for the spin of an electron. The textbooks are full of statements that electron spin cannot be interpreted as the intrinsic moment of physical (i.e. spatial) rotation of the particle. However, just this can be easily done using the Dirac equation, so that the spin of an electron can be interpreted as the moment of purely physical rotation of the wave function [5].

The paradoxes listed above can be also correctly considered to give a thorough understanding of the fundamentals of the modern physics and its limitations in describing the world. Physics is not so all-powerful (even theoretically) as most people (including even many professional physicists) tend to believe. Moreover, these limitations can be derived from the already established principles of physics itself, without inventing new theories. These limitations lead to the paradoxes, which have, despite their apparent dissimilarity, the same source and ground.

We are going to use Nonlinear Dynamics as the instrument for our studies. Nonlinear Dynamics is the science that aims at deeper understanding of the existing fundamental physical laws without trying to invent new ones. Another aim of Nonlinear Dynamics is finding common properties (and corresponding laws) in physical systems apparently different in terms of application region, but characterized, however, by very similar evolution.

**Section 1. The quantum mechanical Schroedinger's "Cat paradox"**

We are not going to give a detailed description of this paradox, because it is thoroughly treated in a number of papers [3, 6, 7, 14]. We shall concentrate on its essence. Cat is a macroscopic system, which, being described in terms of quantum mechanics, i.e. microscopically, can exist simultaneously in the two states, namely, live cat and dead cat. Thus the cat is a "weighted sum" of these states or, in other words, a wave packet of the psi-functions. In fact, both a possible outside observer and the cat itself can only recognize one of these states with probability determined by the squares of the corresponding "weight factors" in the abovementioned "weighted sum". Speaking mathematically, this corresponds to the transition from the mixed to pure state and diagonalization of the density matrix (see Appendices A, B).

This process is called wave packet reduction and it is not described by the equations of quantum mechanics. Moreover, it corresponds to the increase of microscopic entropy (determined via the density matrix). This connects wave packet reduction to the previous two paradoxes and seems to contradict to the reversibility of quantum mechanical equations and to the "returns" in quantum systems. Such paradoxes can be found also in classical mechanics, and we shall prove this by demonstrating further a classical analogue of the Shroedinger's "cat paradox". This analogue does exist in classical mechanics despite the general belief in the "purely quantum mechanical nature" of the "cat paradox". This widespread belief is due to the lack of thorough understanding of the essence and grounds of these paradoxes. The necessity to introduce this mysterious wave packet reduction for macroscopic systems (which are, as a rule, the final objects in the measurement systems, i.e. detectors) lies at the base of the Shroedinger's "cat paradox".

Attempts have been made to describe the states of consciousness corresponding to the quantum "weighted sum" using the means of art [8]. However, even if such states do exist, they are rather exceptional than common.

The apparently simplest solution of the paradox would be that the evolution of the system just bifurcates into branches each characterizes by a certain probability. The analogy can be found in statistical mechanics where the state of the system is described by a "cloud" of points in phase space, and not by just a single point. The matter is that wave packet reduction, in contrast to the "branching evolution" in statistical mechanics, leads to evolution of branches other than it would have been, had the “weighted sum” been preserved. This means that the two possible evolution branches influence each other via mixing of the results of the final measurement (which is, in essence, reduction) performed for the “weighted sum” of these branches. The reason for this is that the solution is a direct sum, and not a statistical sum of the branches.

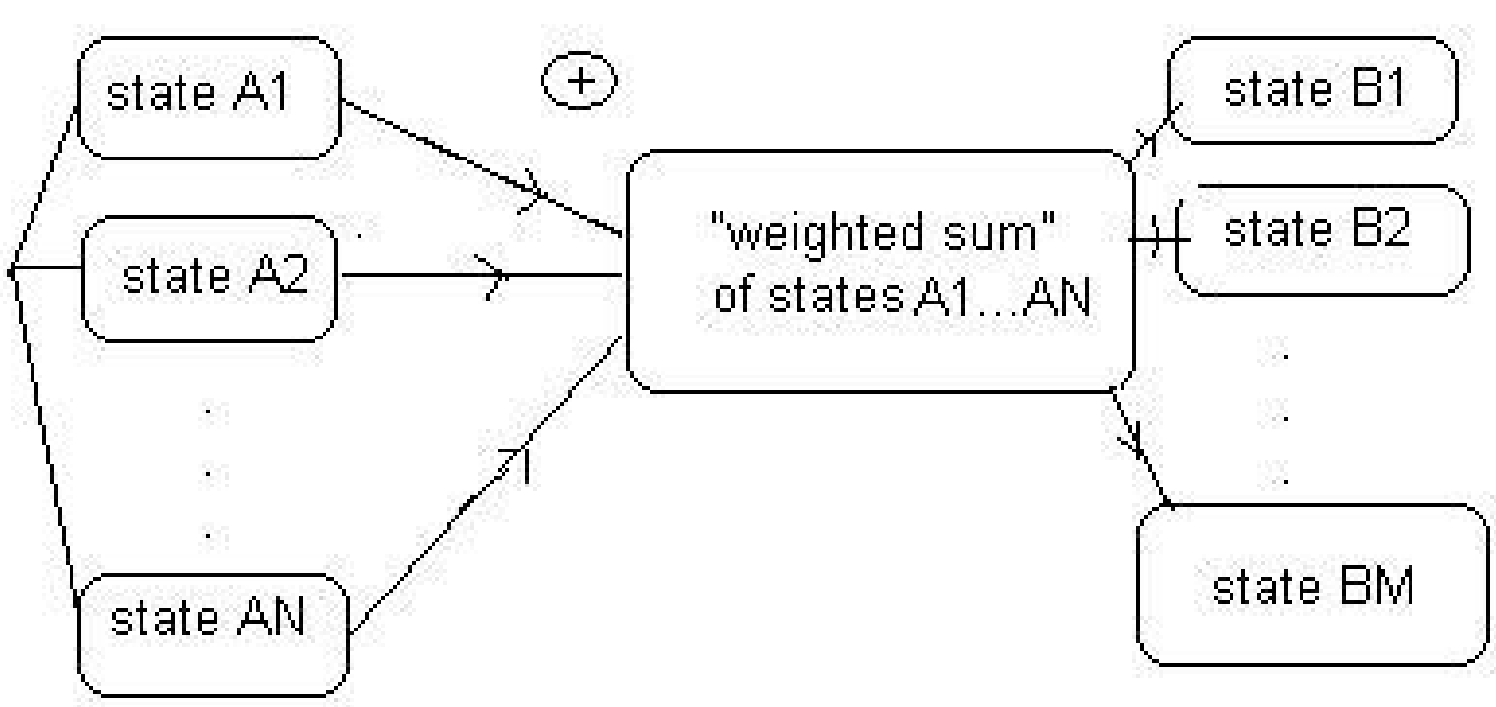

Fugure 2a
"Weighted sum" evolution of states correspondent to different values of macroscopic variable A after measurement of macroscopic variable B and "wave packet" reduction.

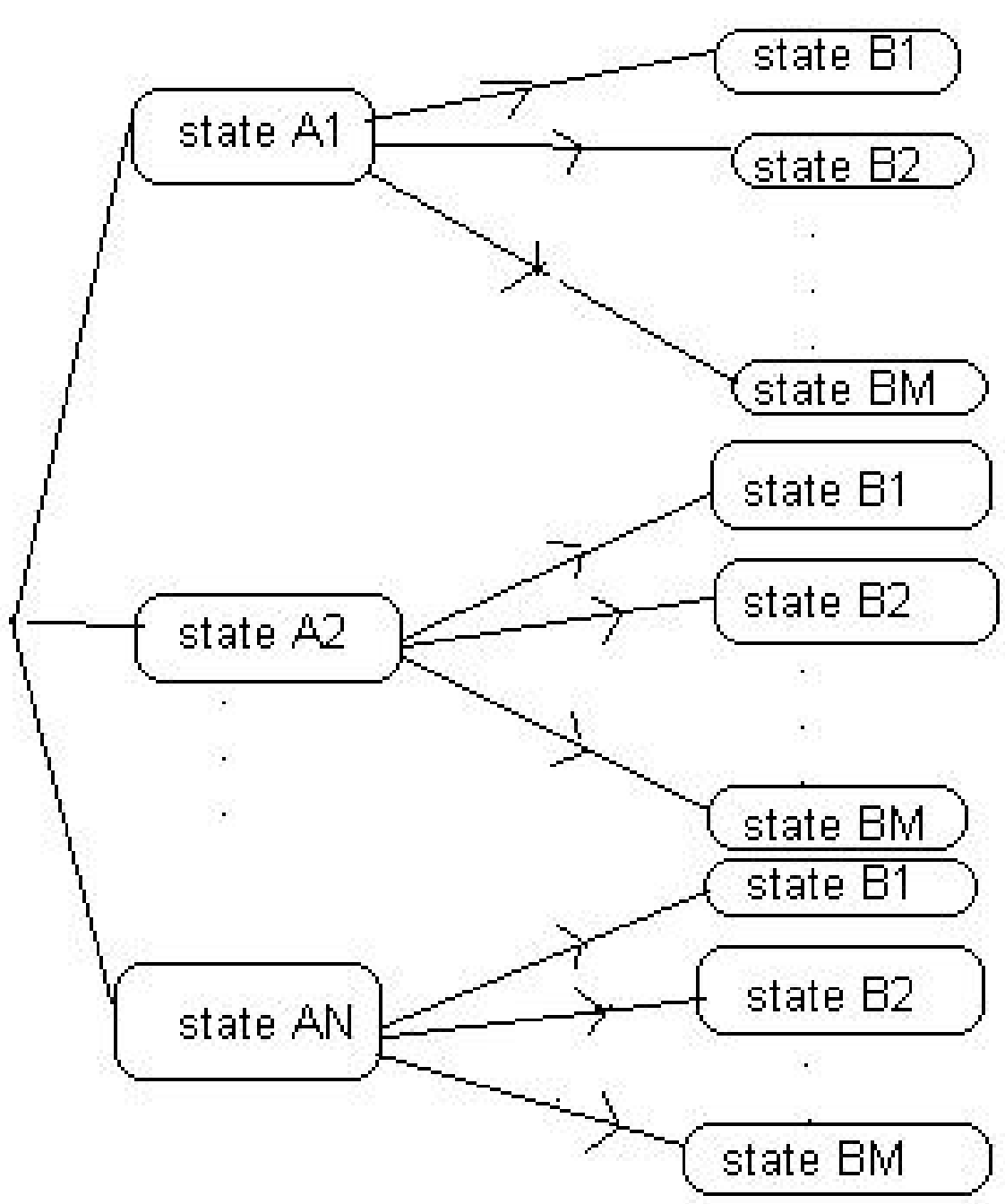

Figure 2b Solution is a statistical sum of the branches, and not "weighted sum" reduction as on Figure 2a.

This thesis is confirmed by the following examples. The first one is connected to the reversibility of the quantum mechanical equations of motion. As it was already mentioned, wave packet reduction leads to an entropy increase, whereas reversible quantum evolution leaves it unchanged. We can hardly expect that these two types of evolution, being so different, can give the same result.

The second example is connected to the Poincare's theorem of the "returns" of a dynamical system to the vicinity of its initial state [1, 2]. Further we discuss some properties of this theorem specific for quantum systems. In classical mechanics most of the dynamic systems are chaotic ones: the return times form a random series and vary strongly even for small variations of the initial conditions. However, a small but important class of systems has periodical or almost periodical return times, which are stable with respect to small variations of initial conditions. These are systems integrable in action-angle variables.

The situation with the quantum systems is quite the opposite. A closed quantum system with limited volume and limited number of particles always has periodical or almost periodical return times. The return times tend to infinity only for systems infinite volume or infinite particle number [9].

Therefore, returning to the "cat paradox", we can conclude that the "reduction of the cat" leads to an increase of microscopic entropy. This makes the return to the initial state characterized by a lower entropy impossible, whereas usual reversible quantum evolution preserves the same entropy value. Thus, reduced and non-reduced systems have different dynamics.

The third interesting example is the "watched-pot-that-never-boils" [3]. Consider a particle that should undergo a transition from a higher level to a lower one according to the laws of quantum mechanics (e.g., decay). Observing the particle too often prevents its ever doing this! Disturbances of the system's evolution caused by the observation leads to this effect. How often do such reduction acts really occur and what is their duration? This question does not have a clear answer.

By the way, this paradox explains why the decay of particles or states in quantum mechanics is not precisely exponential, but only approximately so. These deviations could apparently allow to determine the duration of the system's evolution (its "age"), because purely exponential type of decay makes it impossible. However, this is not so: the perturbations due to reduction during observation and the observation process itself make the difference between the exponential and quantum decay types unobservable.

These considerations prove that reduction is not just a mechanical separation of motion into independent branches. It completely changes all the real dynamics of the system.

Further we consider the known approaches to the solution of the "cat paradox". All these approaches make certain steps in the right direction, but do not lead to a complete solution of the paradox.

Many different interpretations of quantum mechanics appeared recently. Many-world concept [3] is one of the best-known and most popular ones. This interpretation implies that the possible evolution branches exist in simultaneously parallel "worlds", and the reduction does not call for choosing a single branch, because the observer can see only one branch in each "world". However, since all the branches exist at the same time, all the possible "worlds" influence the result of the measurement and thus ensure the interaction of the branches. Such interaction inherent in quantum mechanics is broken in standard reduction by choosing only a single evolution branch. The problem with this interpretation (as with most others) is that it does not solve the difficulty, but removes them to another, less obvious plane. Thus, an observer in each of the possible "worlds" has complete information concerning this "world" only, and has no direct information concerning other "worlds". This makes the available information incomplete and insufficient to predict the measurement results, which are determined by the total evolution in all the "worlds". Thus, disruption of the normal quantum evolution in case of reduction is replaced with the principal impossibility to predict the measurement results due to the limited information in each of the "worlds". However, this concept contains a step towards the correct solution of the paradox.

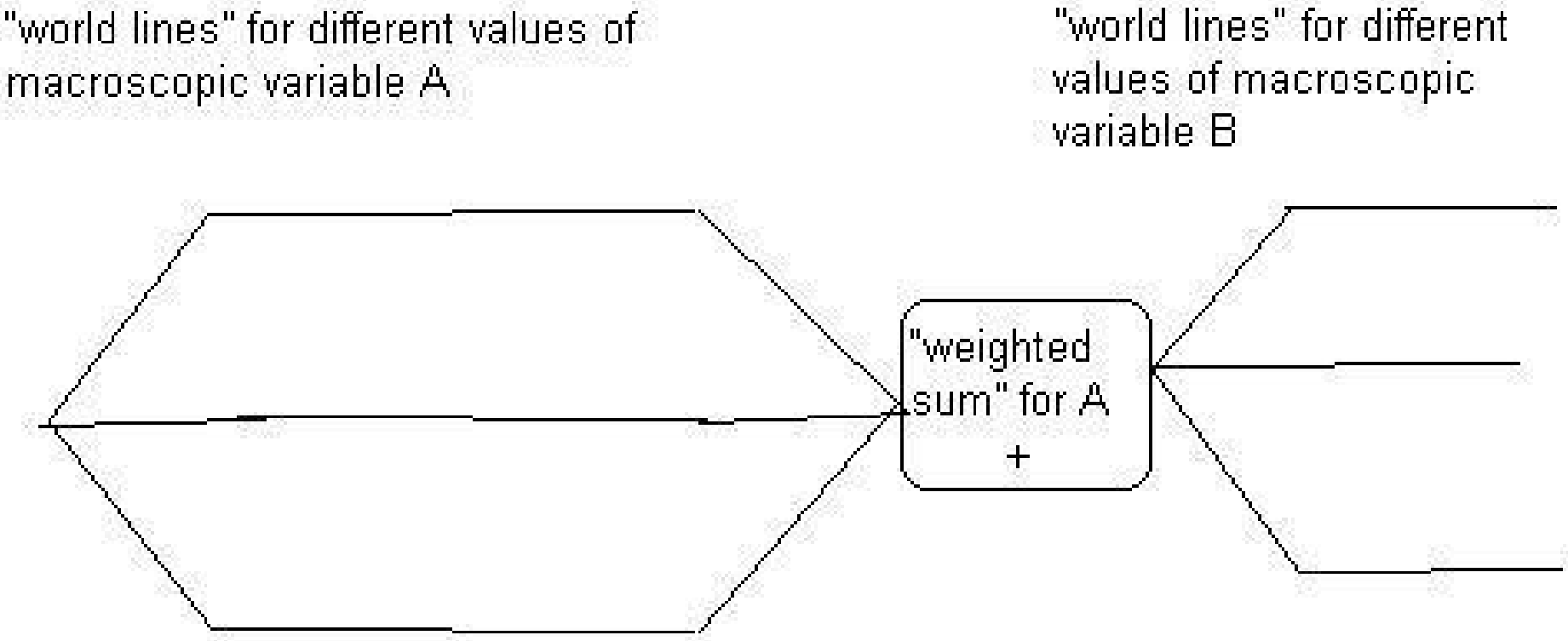

Figure 3 Many-world concept.Possible evolution branches exist in simultaneously parallel "worlds", and the reduction does not call for choosing a single branch, because the observer can see only one branch in each "world".
However, since all the branches exist at the same time, all the possible "worlds" influence the result of the measurement and thus ensure the interaction of the branches. So the problem of such many-world concept is that the observer in any "world" have information only about this "world" and can't predict the result of the measurement, that depends on all "worlds".

The next approach to the solution of the "cat paradox" consists in accounting for the influence of the external observer. Wave packet reduction occurs at the moment of measurement or observation. Thus the reduction and disruption of the normal quantum evolution can be explained in this case by the influence of external forces in the non-isolated system. One of the principal basic differences between the classical and quantum systems should be discussed here. The influence of the measuring device on the measured system in classical mechanics can be made infinitesimal (certainly, only theoretically, and not practically). In quantum mechanics, however, the measurement cannot be performed without wave function reduction and a finite, if small, impact on the measured system. The more precise information about the quantum system we want to obtain, the greater perturbation should we introduce into the system. This makes quantum mechanics essentially incomplete, and this (and not, as generally assumed, different type of basic equations) does make quantum mechanics different from classical mechanics. Note that this does not make impossible either precise reversible quantum mechanical description of the systems or its experimental testing by an external observer. However, the possibility of such description is limited and subject to certain conditions, which shall be discussed later. At present we only note that the external observer can only verify the correctness of the reversible quantum equations of motion themselves, but he cannot test the exact variation of parameters and the history of the system's evolution described by these equations. The matter is that evolution is determined both by the equation type and by the initial conditions for the system, and the latter is necessarily disturbed by the measurement.

The impossibility of excluding the influence of the measurement is illustrated by the experiment with a quantum falling on a screen with two slits and an interference pattern behind the screen. If we determine, which of the two slits the quantum passes through, then the influence of the other slit becomes impossible (due to the causality principle and limited speed of light), and the interference pattern disappears, replaced by the sum of intensities of the signals from the two slits.

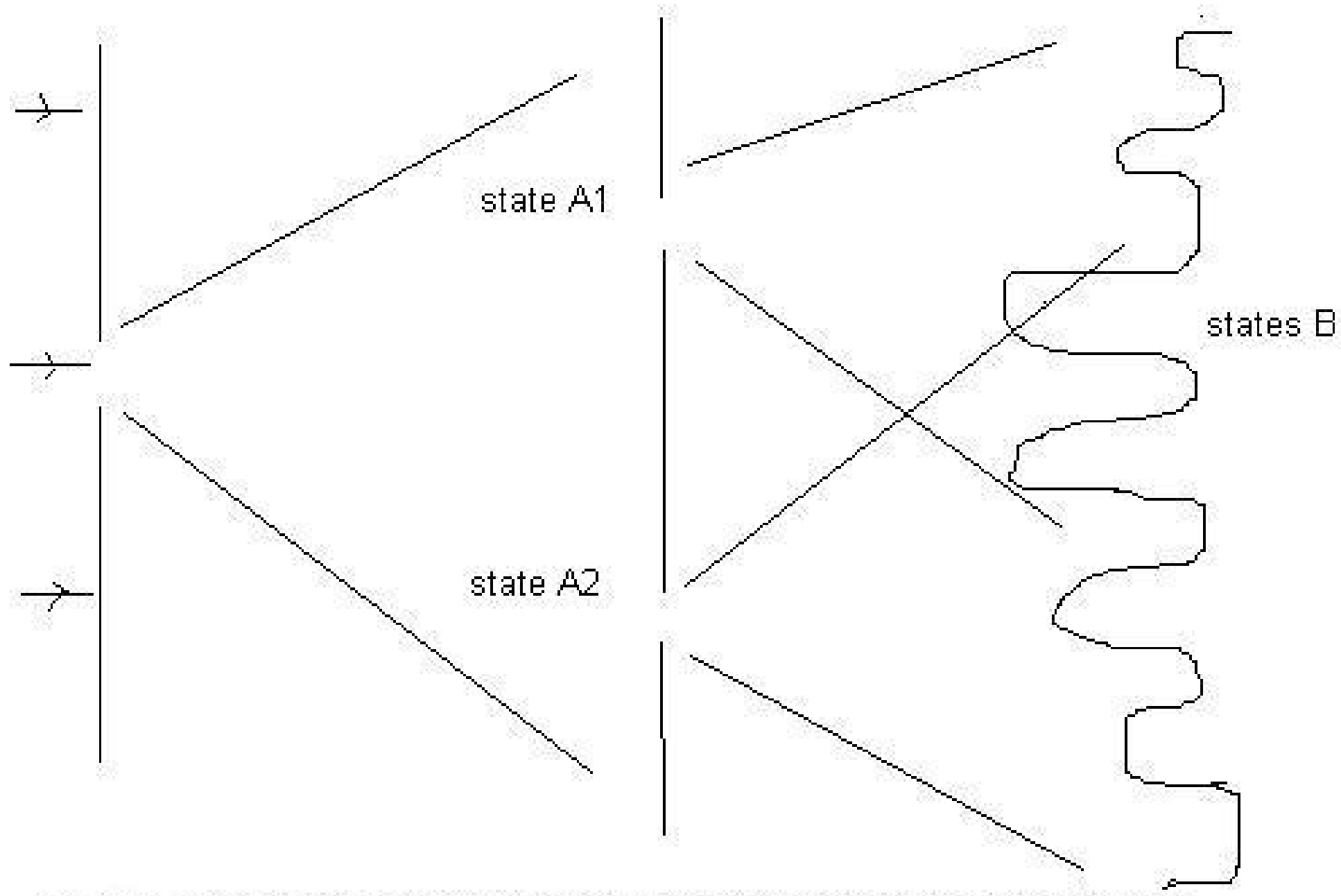

Fifure4a interference pattern from the two slits in the case of no detection, , which of the two slits the quantum passes through
example of the case on Fugure 2a

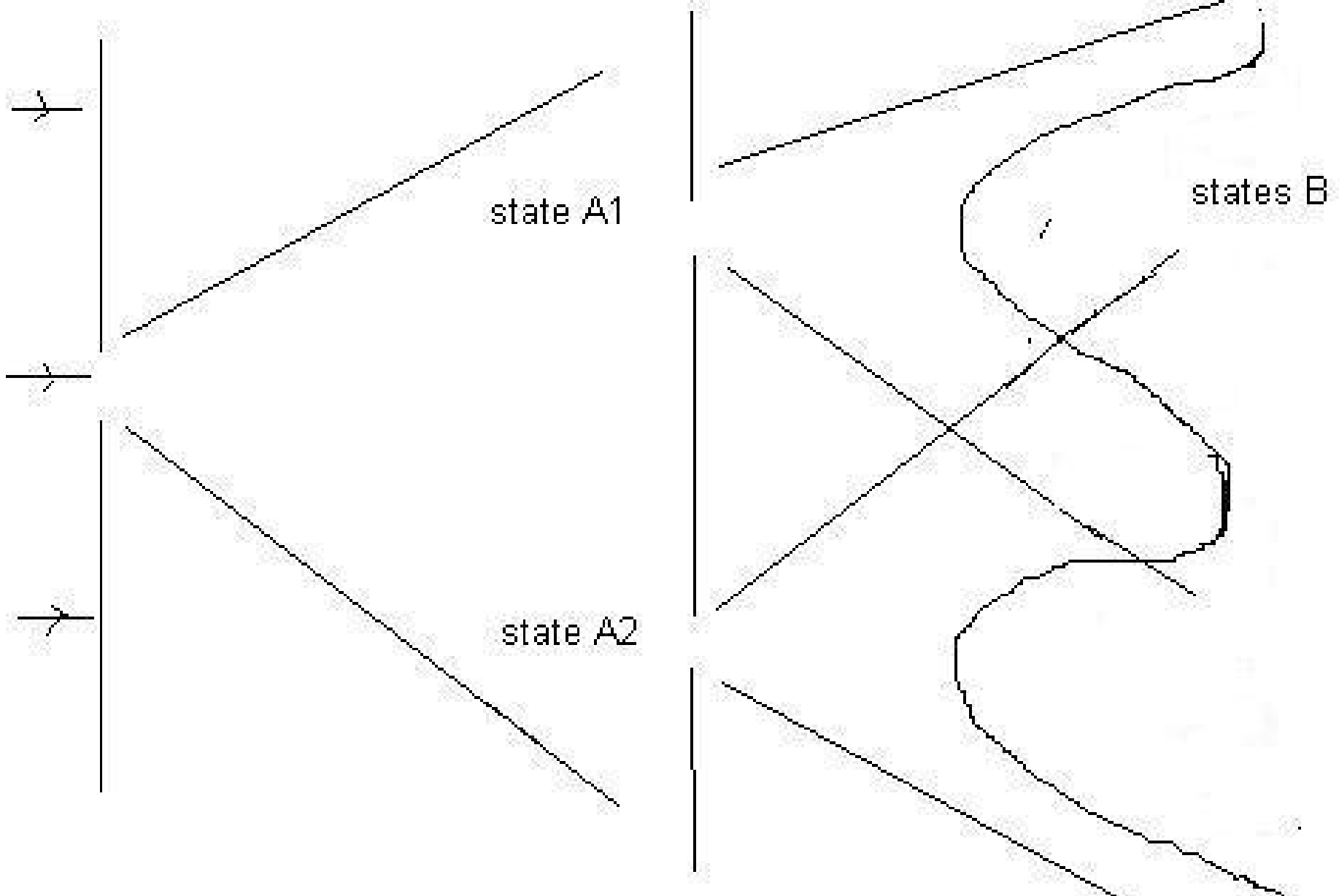

Figure 4b The sum of intensities of the signals from the two slits in the case ot detection, which of the two slits the quantum passes through
Example of the case on Figure 2b

In order to make the process of the observer interacting with the system more objective and independent on the subjective observer, this interaction can be replaced by a non-controllable interaction with the macroscopic environment. This process called decoherentization [16], which makes the quantum system non-isolated, is deservedly popular. Its main advantage is in replacing the set of reversible quantum mechanical equation of motion for an isolated system and the mysterious irreversible reduction process with a closed set of equations including an external noise. These equations have a definite and purely practical meaning. However, this implies new difficulties. The less essential one is that the parameters of the system and the equation become dependent on the choice of the type of a certain non-controllable external influence. What should this choice be in order to ensure a correct description of the system?

The far more essential difficulty is that both the observer and the environment can be, in principle, included into the system under consideration. Then the reduction process, which is quite real and cannot disappear in this case, will occur in a closed system. But then the abovementioned explanation of reduction as a result of a certain external influence is invalid, and we return to the paradox in its previous unresolved form. The simplest example clarifying this phenomenon is the case when the cat is self-contemplating, determining that it is still alive, and not dead. Then what is there to do? Should we introduce an infinite chain of consecutive observers thus expanding our system to infinity? Should we attribute a real material force to our consciousness, and the ability to reduce the system's wave function? Or, as it is sometimes suggested, create a new "physics of consciousness" [7]? The true solution is much more simple and beautiful.

In case when the cat is under external observation and thus neither the observer nor environment are included into the system, there is no paradox, as we could see. The problem consists in clearly determining the exact possible type of external influence, which is definitely impossible, because the external factors are not included into the consideration. What should be done if all the external factors are included in the system (which is also the case of the cat himself being the observer)? Evolution and dynamics of the system would be essentially different in case of reversible dynamics without reduction, and the irreversible dynamics with reduction. This is most clearly illustrated by the absence of returns of the system to the initial state in case of reduction (due to the entropy increase in this process) and the almost periodical returns predicted by the reversible equations of quantum mechanics. The solution of this paradox consists in the following: the difference between these two types of dynamics, though quite real, cannot be verified experimentally. Indeed, since a closed isolated system cannot either measure its own state precisely and unambiguously, or solve the set of equations describing its dynamics. Neither can it perform a complete experimental verification of its own dynamics, nor make an unambiguous choice between the two possible types of dynamics, because in the considered case both of them lead to the same result in the region under test. Thus, the system cannot verify the Poincare's theorem concerning its own return to the initial state. In fact, the system should somehow "remember" the initial state in order to perform this verification by comparing the initial and the current states. However, a complete return to the initial state makes the existence of such "memory" impossible. In fact, such "memory", being a part of the system, should also return to the initial state when the latter does, thus the "memory" will be erased. Thus, on return to the initial state, the system should inevitably loose the memory of its whole history and cannot register the return. Only an external independent observer is capable of performing such verification. On the other hand, the observer (or the environment) should inevitably introduce a disturbance into the quantum mechanical system thus causing reduction. Thus, the possibility of two types of dynamics for the same system is explained by the impossibility of experimentally detecting the difference between the two, and not by some mysterious productive properties of consciousness.

An important note: the ground for this paradox and the reason of its clear manifestation in quantum mechanics is that just in quantum mechanics the measurement implies a finite, if small, disturbance of the measured system. This disturbance, and not the probabilistic nature of quantum mechanics makes the latter so much different from classical mechanics and leads to many difficult paradoxes.

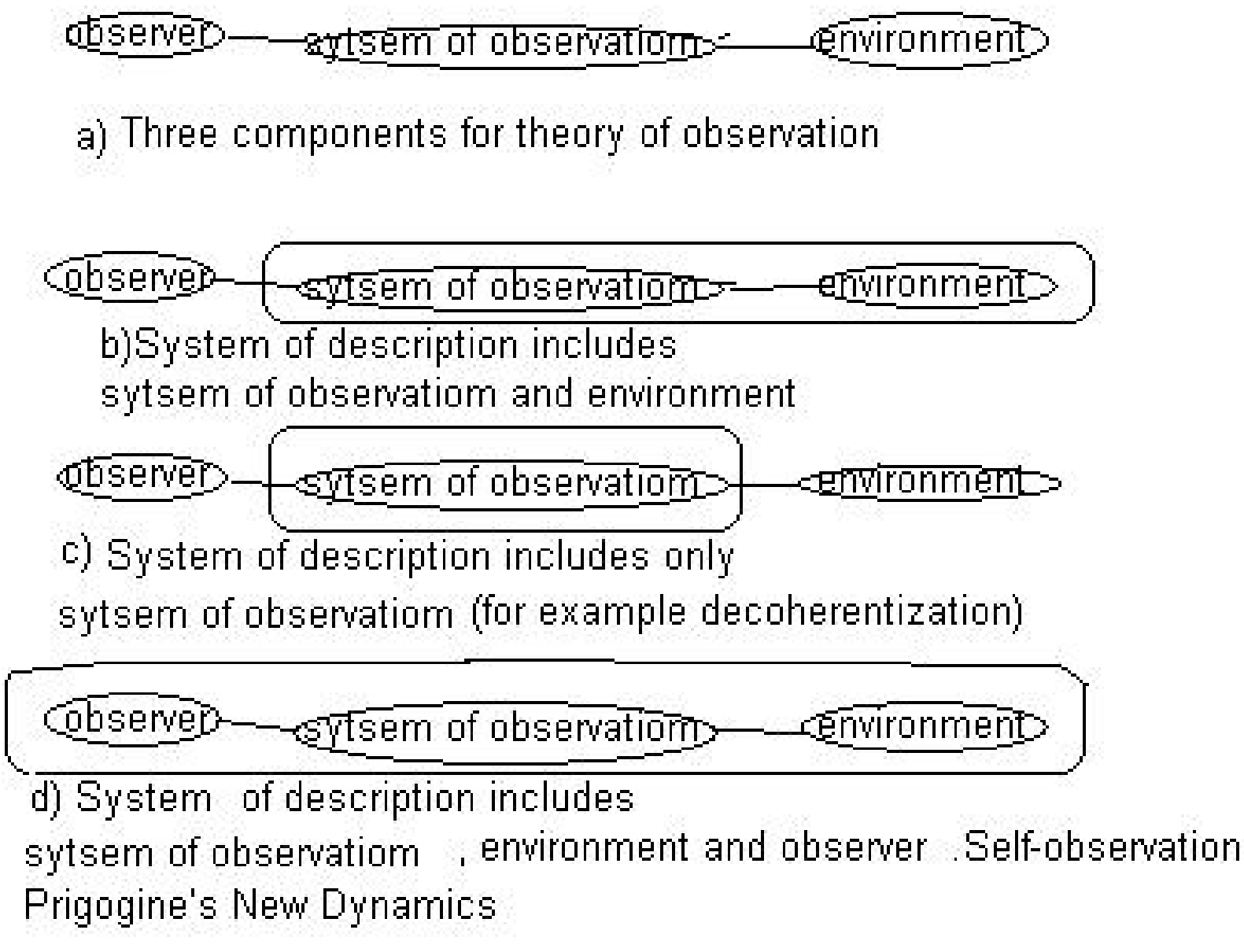

Figure 5 Interactions of three components for theory of observation.

Here we would like to recall that under certain strict conditions (which are discussed below) a precise verification of the reversible quantum mechanical laws is possible, despite the abovementioned limitations.

**Section 2. Classical analogue of the quantum mechanical Shroedinger's "cat paradox"**

The Schroedinger's "cat paradox" is usually considered as purely quantum phenomenon that has no analogue in classical mechanics. However, this is a mistake. An analogy to reduction does exist in classical mechanics. Indeed, irreversible entropy increase (which in quantum mechanics occurs at the moment of wave packet reduction), predicted by the corresponding law, also occurs in classical statistical mechanics. And, just as in quantum mechanics, this increase contradicts the reversibility of the laws of motion and the Poincare's theorem of the system's return to the infinitesimal vicinity of its initial state. The classical laws of motion leave entropy constant if it is determined via density function in phase space. Being defined via the "coarsened" density function (i.e. averaged over a certain vicinity of each point of the phase space), entropy can both increase and decrease. What then explains the experimentally observed entropy increase? Introduction of the Boltzmann's "molecular chaos hypothesis" [1, 2], which is one of the types of "coarsening" the density function in phase space, does. This hypothesis implies the loss of correlations between the velocities and positions of different molecules, i.e. the loss of reversibility of motion in case of velocity reversal. This contradicts the Poincare's theorem of the system's returning to the small vicinity of the initial state, i.e. the evolution of the system becomes irreversible.

Here we can see a perfect analogy to the wave packet reduction. Boltzmann's hypothesis leads to the correlation loss ( these correlations similar to the non-diagonal elements of the density matrix in QM) and irreversibility of motion, just as wave packet reduction does. Thus the "molecular chaos" hypothesis or other types of "coarsening" the density function in phase space is a perfect analogy to the wave packet reduction in quantum mechanics.

What calls for introduction of these additional assumptions in fact contradicting classical mechanics? The reasons for this are the same as for introducing reduction in quantum mechanics: indiscernibility of the two types of dynamics (with "coarsening" and without it) in real experiments.

In closed systems it is due to the impossibility of "memorizing" the initial conditions in self-observation because of the returns, and in case of external observer it is due to the interaction of the system with the observer or with the environment.

However, here we encounter the most essential difference of classical and quantum mechanics. The interaction between the observer and the system under observation is essential for quantum mechanics, and it is finite, whereas in classical mechanics such interaction can be theoretically reduced to zero. In fact, even in classical mechanics, such interaction is always there and always finite. This explains the contradiction between the theoretical possibility of entropy decrease and experimentally observed entropy increase in large systems. The real and finite, if small, interaction with the observer or "environment" leads to the disruption of the processes involving entropy decrease. Indeed, such processes, in contrast to the ones involving entropy increase, are very unstable to the chaotic external influence. This leads to the disruption of such processes and synchronization of the arrows of time between the observer, observed system, and environment even for a small value of the interaction.

The arrow of time is defined in the direction of entropy increase. Introduction of the arrows of time was previously used to explain the law of entropy increase. However, this explanation gave rise to a relevant question: since both direction of the arrow of time have equal possibilities (maximum possibility corresponds to the equilibrium states that do not have a clearly defined direction), then why do all these arrows have the same direction? This was considered an insoluble mystery, and its solution was supposed to lie in the depths of the origins of the Universe [28]. However, the solution is in fact very simple and consists in a real, if small, interaction of all the subsystems leading to the general synchronization of the arrows of time.

Note that the theoretically possible zero interaction between the systems in classical mechanics was the reason why the Schroedinger's "cat paradox" was first so clearly formulated only in the frame of quantum dynamics. In fact, such paradoxes can be found also in classical mechanics in the assumption of a finite, if small, interaction between the systems. This interaction is always there in real life with the exception of several very subtle artificial cases specially set up in laboratory experiments. Further we shall discuss several such cases. First we introduce the following convention: here and below the system is called "real" if it implies a finite if small interaction with the observer or environment (note once again, however, that this interaction can be zero in classical mechanics. )

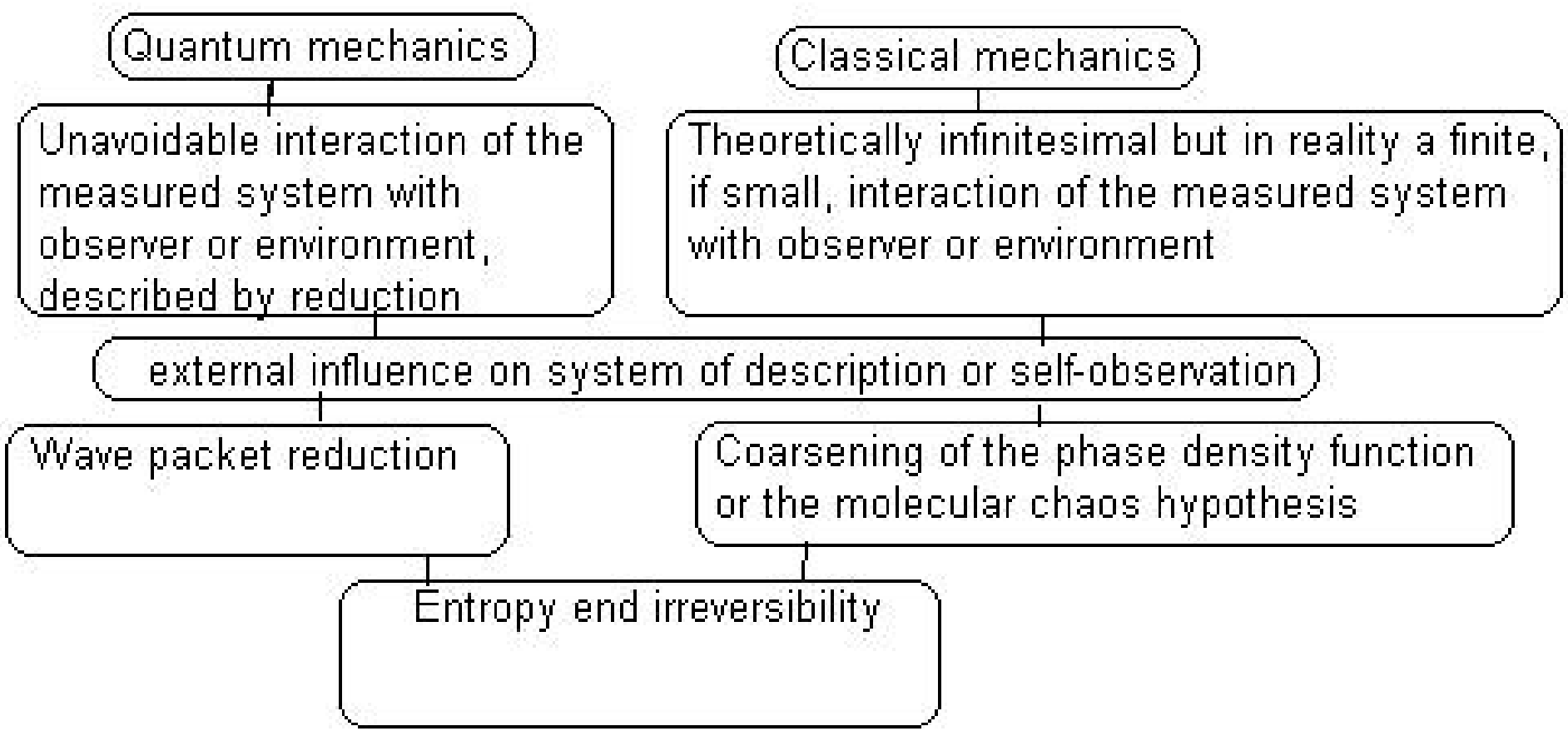

Figure 6 Sources of irreversibility and entropy in physics.

**Section 3. Experimental observation of the quantum Schroediger's "cat paradox» and experimental verification of Ideal Dynamics.**

In this Section we discuss mainly finite size systems. For infinite size system, for example for stars in infinite Universe, we can obtain information about stars by observation of the light from these stars. Such measurement doesn't influence on stars, because the light doesn't come back to stars after observation for infinite Universe.

The ideal system, i.e. the system with (almost) zero interaction with the environment and the observer is easy to set up in classical mechanics. In quantum mechanics it is impossible: the measurement of the system is always connected with an unavoidable interaction. The difference between the reduced and non-reduced types of dynamics might appear to be experimentally unverifiable. (Further we call the non-reduced dynamics Ideal dynamics. In case of classical mechanics this corresponds to the dynamics before the "coarsening" procedure.) However, this conclusion is untrue. There are two possible types of measurement in quantum mechanics, corresponding to the state of the system before and after the measurement with an almost 100% precision. Both the states cannot be measured simultaneously because of the interaction between the observer and the system. We call "preparation" the procedure that measures the state of the system after the measurement, because the initial state is changed and remains unknown. The term "observation" is used in the discussion below to designate the procedure that measures the state of the system before the measurement. Thus the quantum system set into the initial state via the "preparation", almost completely isolated (from environment and measurement) in the intermediate state and measured in its final state using the "observation" procedure is a test case for verifying the Ideal dynamics. Such system is a quantum mechanical analogue of a classical isolated system.

This scheme has the following essential drawbacks:

1) The initial state (before "preparation") is changed and remains unknown.
2) The intermediate states are not measured and remain unknown. It is possible in fact to compare only the initial and the final state.

3) The cases that allow reaching such a complete isolation of the system are rare and require great pains to achieve.

The examples of such systems are the following:

1) Mesoscopic system at low temperatures [10]. Such systems are close to the applicability limits of the law of large numbers due to their large size, and thus can be considered almost macroscopical. The size of quantum wave packets at low temperature (and therefore small molecule velocity and momentum) is very large in accordance with the uncertainty relation. It is therefore close to the system's size and thus the wave packets support quantum correlations. The interaction with the environment is weak and its value is easily controllable. This enables verification of quantum coherent oscillations and tunnel effect in relatively large, almost macroscopic systems. All the experiments known so far confirm that Ideal dynamics is realized, and there is no reduction in the intermediate states.

2) Systems in the vicinity of the second type phase transitions. Correlation length for such systems is large and almost equal to the system's size.

3) Certain types of processes in live systems or primitive prototypes of such processes.

**Section 4. Definitions of Real Dynamics, Ideal dynamics, Unpredictable Dynamics, and Macroscopic State. Prigogine's "New Dynamics" (see Appendix F).**

We agreed upon calling Ideal Dynamics such evolution of the system, which is correctly described by the basic equations of quantum and classical mechanics.

In reality, aside from a very small number of cases mentioned above, Ideal Dynamics is experimentally unverifiable due to the impossibility of the system's complete self-description and since the interaction between the system and environment (observer). This makes the behavior of the system unpredictable, thus giving rise to Unpredictable Dynamics instead of Ideal Dynamics. In practice, however, most of the systems can be well described and their behavior can be well predicted by the laws of physics. Why is this possible?

There are two basic factors leading to the unpredictability.

1) Impossibility of a complete self-description if the observer and the environment are included in the system to be described. This limits the possibility to precisely determine the initial conditions of motion.
2) Non-controllable interaction of the external observer and environment with the system to be described. This puts a limitation on the exact knowledge of equations of motion due to the non-controllable external noise.

However, these difficulties can be resolved. The solution consists in substitution of the complete description of the system with a reduced one via introduction of macroscopic variables being certain functions of microscopic variables. Here we use a very wide interpretation of macrovariables. This means that, e.g., knowing the positions and velocities of all the molecules with finite (however high) precision is also considered macroscopic description of the system.

Remarkable is the fact that most real systems have at least one (often more) set of macrovariables that makes the equations of motion for these systems independent on the external noise or the errors in the initial conditions (over a wide range of their values and types). (This should be understood as independent on almost independent on the value and type of this noise or the errors in the initial conditions (over a wide range of their values or types) during a period of time less than half the return time for periodical or almost periodical systems or even infinite time for chaotic systems or systems with an infinite number of particles or infinite size).

Thus, the type of the system's evolution that we refer to as Real Dynamics does not depend on the errors or external noise, and depends purely on the properties of the system, just as Ideal Dynamics does. There are at least to grounds that make Real Dynamics stable against noise: the statistical law of averages and discreteness of quantum transitions ensuring the stability of chemical bonds [11].

Here we encounter the very important issue of choosing the macrovariables. The requirement of independence of the system's evolution on noise limits the possible choice of macrovariables. E.g., instead of the live-dead pair of states of the cat in Schroedinger's paradox we can choose in quantum mechanics the half sum and half difference of these states to be our new basic set. Why in reality the choice is always for the live cat - dead cat pair of states? Because this pair is stable against the small noise introduced by the macroscopic environment, whereas states corresponding to their half sum and half difference decay even in case of slight external noise (in accordance with the Daneri-Loinger-Prosperi theorem [14]). Other limitations can also be imposed on the macrovariables in order to reduce their number or to make their behavior more deterministic.

Another important property of Real Dynamics is the ambiguity in the choice of the type of this dynamics and the set of macrovariables it describes. In fact, Real Dynamics implies developing new fundamental laws for this description level based only on the exact Ideal Dynamics, which becomes experimentally unverifiable. However, the exact and final choice of these laws is mainly determined by convenience considerations.

All this allows substituting the Unpredictable Dynamics with the predictable Real Dynamics formulated in terms of macroscopic variables. Real Dynamics is derived via introduction of "coarsening" or "reduction" or many other similar methods. We would like to note here that in most cases the Real Dynamics is not just a certain approximation of Ideal Dynamics (as it is often interpreted). There is in fact a real difference between the two types of dynamics, but it is NOT observable experimentally. Thus the question which of these types of dynamics is more true is irrelevant.

The paradox of "kettle that never boils" is applicable for the case of particle decay or transition from one quantum mechanical state to another. This paradox is following: if time intervals between the moments of observation of a certain transition (decay) are small enough, then the transition (decay) can never occur. Wave packet reduction occurring at the moment of measurement changes the normal evolution of the system if the measurements are performed too often. However, if the time interval between the measurements is large enough (i.e. much greater than $t = E/\hbar$, where E is the energy difference between the levels, and $\hbar$ is the Planck constant), then the decay process is not disturbed, and is almost independent on the time interval between the measurements. This is in fact the domain of "Real Dynamics".

A Real Dynamics can be attributed to one of the following two types corresponding to the two unpredictability factors:

1) The observer and the environment are included into the system to be described. This limits the exact knowledge of initial conditions of motion and leads to the Real Dynamics independent on the initial conditions over a wide range of their values. This is the most popular type of Real Dynamics, because it appears to be "more objective" and to be independent on the external factors (whereas, in fact, both types of Real Dynamics are determined only by the parameters of the system). The wide known "New Dynamics" developed by I. Prigogine [9, 12] belongs to this type.
2) The observer and the environment are NOT included into the system to be described. The non-controlled interaction of the external observer and environment with the system to be described limits the exact knowledge of the equations of motion. Thus the non-controllable external noise leads to the Real Dynamics independent on this noise over a wide range of its value and type. This case corresponds to the widely used and applied in quantum mechanics “decoherenization” of quantum systems interacting with “large” external macrosystems.

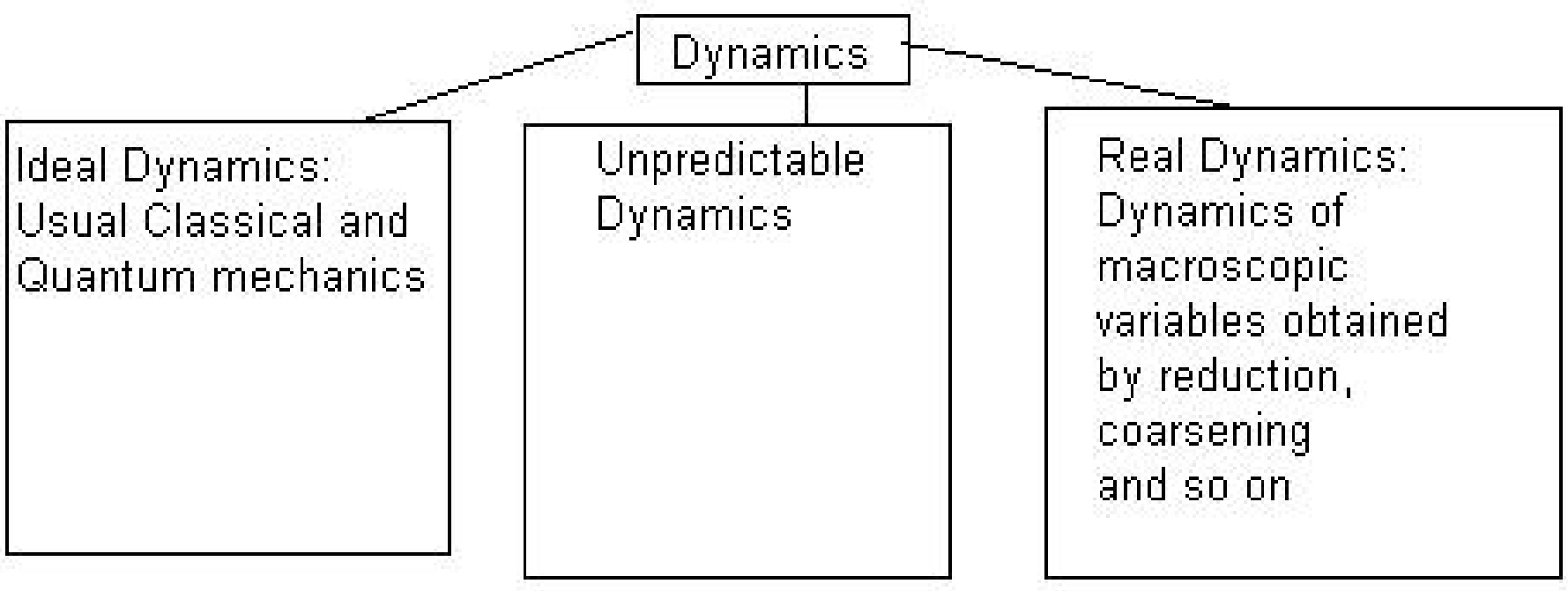

Figure 7 Three types of Dynamics.

Further we discuss the “New Dynamics” developed by I. Prigogine [9, 12] (see Appendix F) in more detail. This type of dynamics differs from other similar techniques due to a very convenient “coarsening” procedure. The matter is that most of the real closed systems in classical mechanics are the system with mixing, where most trajectories are exponentially unstable. Their analogues in quantum mechanics are systems of infinite size or systems with infinite number of particles. Prigogine’s theory was developed to describe just such systems. This leaves open the issue of describing an important if limited class of periodical or almost periodical classical systems and almost all the closed quantum systems, which almost always have the same periodicity properties. The quantum and classical systems described by the Prigogine’s theory are essentially different in this respect. The infinitely large quantum systems have infinite return times, whereas returns times of the classical chaotic systems are finite if arbitrary. However, due to the inevitable errors of self-measurement in classical systems the initial conditions are spread over a small vicinity of the initial point. Therefore the state of the system can be considered to be represented by this small vicinity, and the complete return of the system can be considered as return not just the single point but to the small vicinity of this point. And since the return times are arbitrary, then time of such complete return can be only infinitely large. The arbitrary return times are not self-observable in systems represented by a single point in phase space, and a real observer always introduces an error modifying the final result.

Phase density function has the property of preserving the phase volume of the initial small vicinity. Since initially close trajectories in systems with mixing diverge exponentially in one direction, they should converge in the other direction with the same rate. The “coarsening” is suggested to be made in the latter direction, i.e. in the direction of trajectory convergence. The maximal extent of this coarsening is determined by the requirement of independence (or a slight dependence) of macrovariables on the value of this coarsening, and this relation should be used in Real Dynamics. This coarsening procedure has a remarkable property, which makes it different from many other possible procedures. The equations of motion for the “coarsened” and “not coarsened” phase density function remain equivalent in this procedure. This means that solving the equations of motion is permutable with the coarsening procedure. Thus, either the coarsening can be performed first and then Prigogine’s equations can be solved to obtain the final coarsened phase density function, or the equations of Ideal Dynamics can be solved and then the coarsening be performed.

Velocity reversal does not change the area limited by the non-coarsened density function in phase space. Most of the usually applied coarsening techniques preserve this property, so only the reversibility of the equations of motion is broken. However, as discussed above, this reversibility cannot be observed experimentally in real systems. The irreversibility of the Prigogine’s equations is caused by the asymmetry of coarsening procedure with the respect to the velocity reversal.

The initial state of the system in the phase space of coordinates and velocities of all the molecules corresponds to a compact “droplet”. The conservation law for the phase space has the following implication. Spreading in one direction due to the exponential instability (the “mixing” condition) is accompanied by squeezing in the orthogonal direction, so that the “droplet” is transformed into long and thin “branches”. The direction of these branches corresponds to the direction of exponential spreading. The concept of “coarsening” Prigogine suggested consists in averaging (“coarsening”) the phase density function only in the direction of “squeezing”, and not in all the directions. Widening of the phase density function in the direction of “squeezing” has almost no impact on the dynamics of this function. In case of inversion of velocities (i.e. inversion of time) the direction of squeezing becomes the direction of spreading, and the dynamics of the function, obtained from coarsened function by inversion of velocities, shall differ strongly from the dynamics of the function, obtained from NOT coarsened function by inversion of velocities. Thus the desired “asymmetry” of time is introduced.

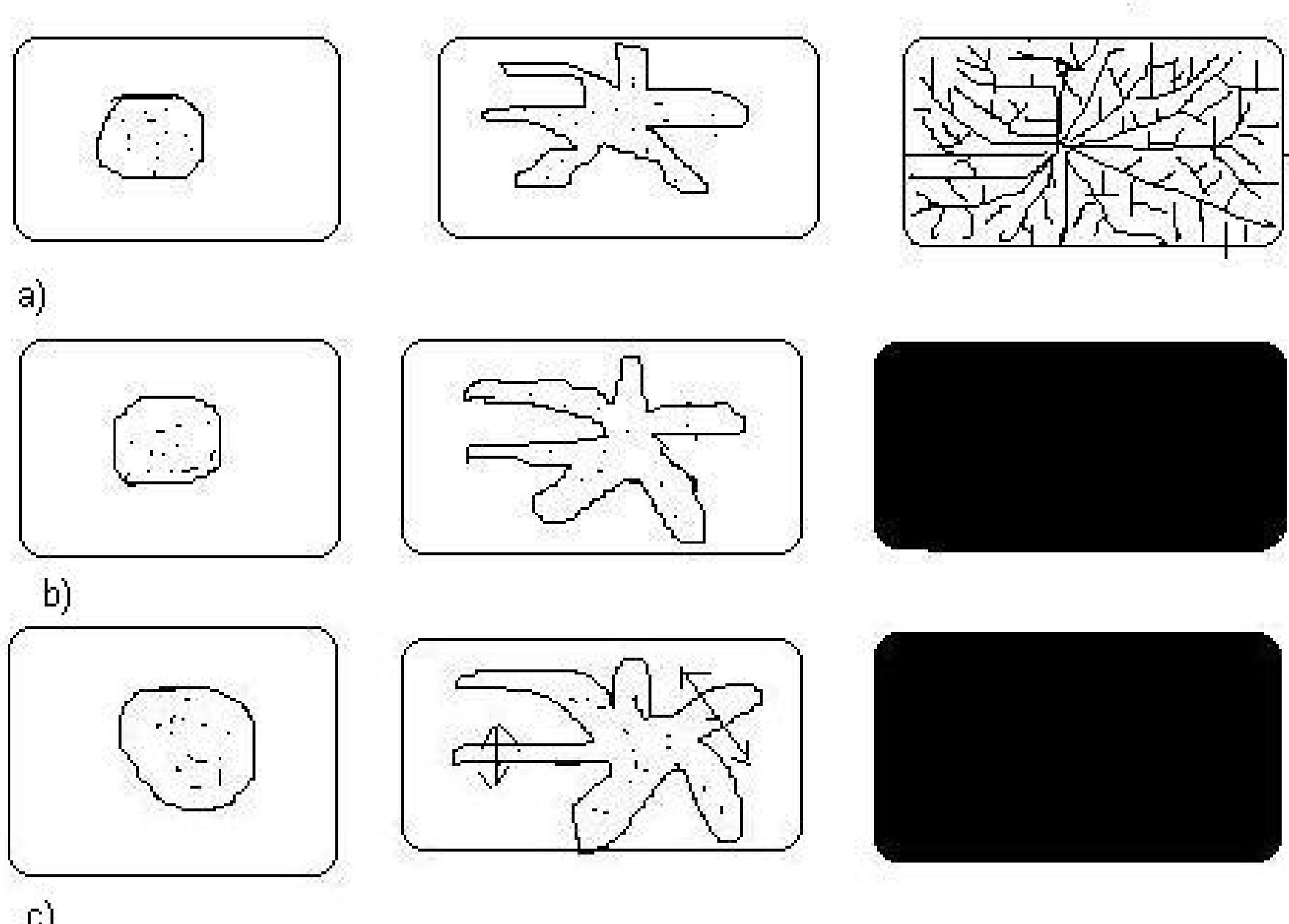

Figure 8a a)phase space "droplet" spreading. b)with isotropic "coarsening" c)with anisotropic Prigogine "coarsening" .<-> is direction of "coarsening"

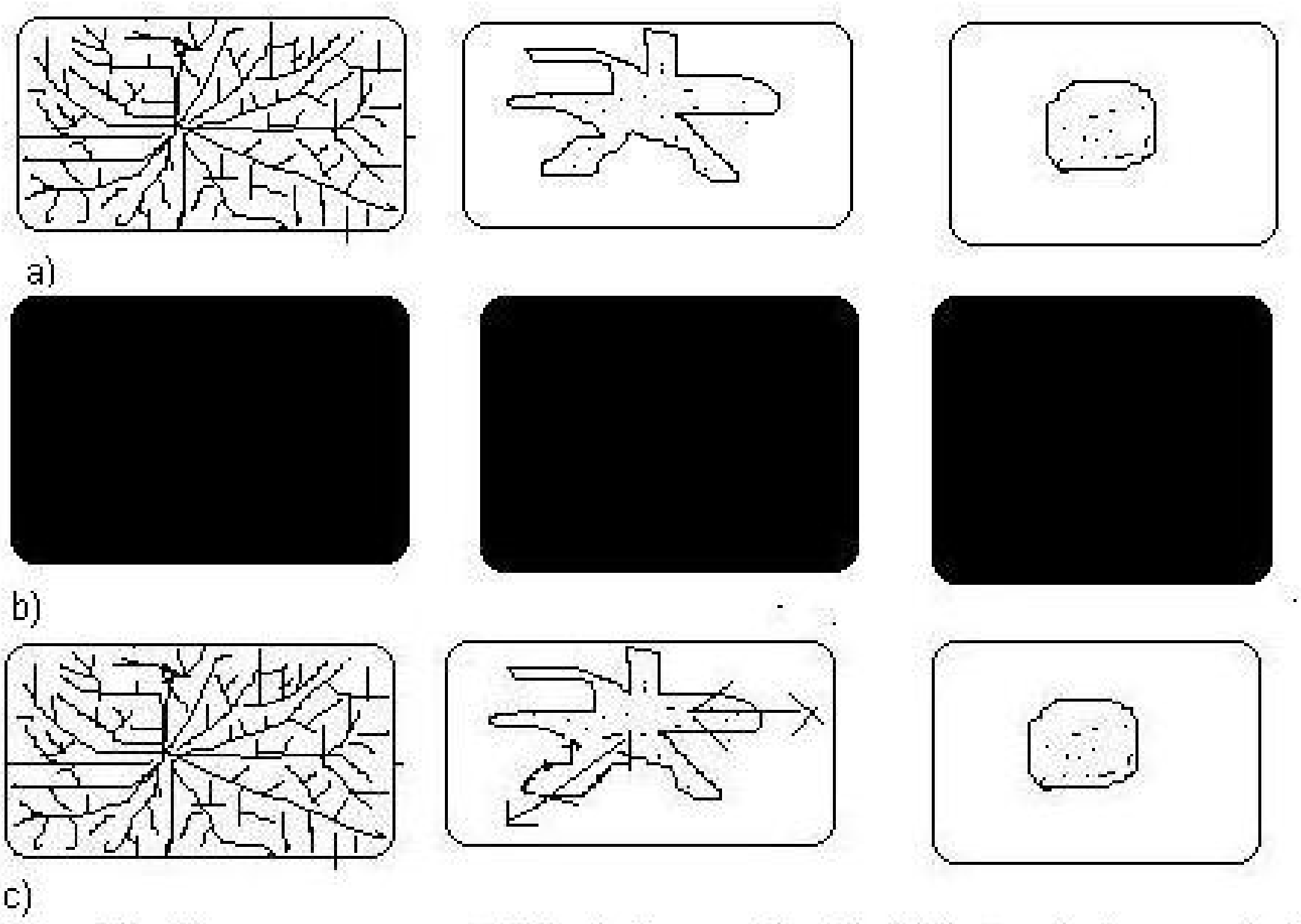

Figure 8b a)inverse process to "droplet" spreading b)with isotropic "coarsening" c)with anisotropic Prigogine "coarsening" <-> is direction of "coarsening" anisotropic Prigogine "coarsening" almost doesn't change dynamics.

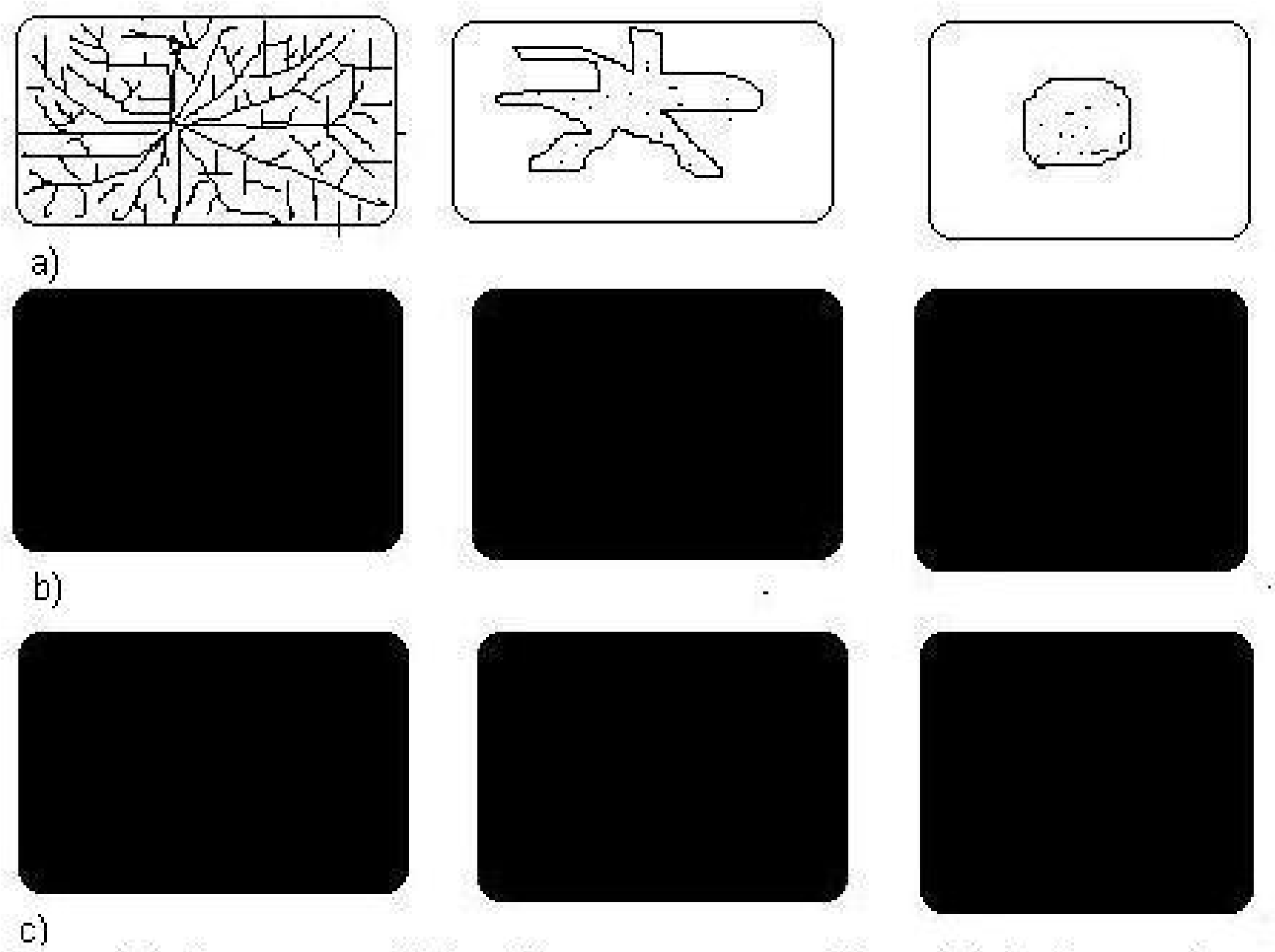

Figure 8c Processes obtained from processes on Figure 8a by inverse of molecules velocities in final state.Processes with "coarsening" are irreversible.

The entropy of phase density function increases with the increase of the phase volume covered with phase "droplet". Hence, as it is possible to see from Figures, for not coarsened functions the entropy does not changes, and for coarsened increases. However, thus for the anisotropic Prigogine coarsening dynamics changes much less, than for the isotropic, usual coarsening.

What the extent of coarsening should be? It should be chosen so as to ensure real experimental indistinguishability of the "New" Real Dynamics and the Ideal Dynamics over infinite time for the above-mentioned systems with mixing. This condition was not discussed in the papers of the Prigogine's school, either.

A similar coarsening can be performed for periodic and almost periodical systems, but in this case the Ideal and "New" Dynamics shall be indistinguishable only during half of the return time. This is in fact quite sufficient for the description of the system, since such returns in real systems are unobservable experimentally, as it was shown before.

The question is much discussed whether the Ideal Dynamics or the Real "New Dynamics" introduced by Prigogine is true [28]. This discussion is similar to that whether Earth goes around the Sun or *vice versa*. In fact, by the very definition of motion the issue is left open and the choice can be made based on beauty and convenience considerations. A similar situation can be found in mathematics, where the choice of definitions and theorems is determined by the same considerations. The theory of this choice is yet an unexplored continent in the foundations of mathematical science, in contrast to the Goedel's theorem. **(A note concerning the Fermat's theorem. The difficulties in proving the Fermat's theorem were often explained by the result of Goedel's theorem. The final conclusion was very close to this statement. Fermat's theorem was deduced as a consequence of a much more general theorem [13], which involves much more axioms than the natural numbers theory that allows formulating Fermat's theorem. In this context it would be extremely interesting to have the mathematicians present not only the 150-page long proof of the theorem, but also a complete (and a quite limited) list of mathematical axioms showing which of them were used in proving this theorem, and which were not. Similarly, proving the consistency of arithmetics involves introduction of transfinite induction and corresponding extension of the list of axioms. However, there is no general theorem in this case, and the new axiom of transfinite induction is introduced "by hand". What statement should be considered axiom, and what – theorem is a matter of beauty and convenience, after all.)**.

Similarly, the difference between the "New" and Ideal Dynamics is experimentally unobservable in most of the real cases, so the choice between the two is mainly free. In those rare cases when the Ideal Dynamics can be verified exactly, it was so far proven to be true. Perhaps, the studied systems are not large enough? Only further experiments can give an answer to this question. We can only guess what type of Dynamics would prove the truest among many. The case is similar to the issue of verifying the Great String Theories and Grand Unification theories: most of these are purely hypothetical, and centuries might pass before an experiment can be set up to give an answer, unless Cosmos or little green men help us. On the other hand, the Einstein's theory of gravity is considered indubitable truth (perhaps, by virtue of habit), though its experimental verification is still very far from complete (**remember the theories of gravity developed by Milgrom and Logunov and the mysterious dark matter**). Nobody knows how much time would pass before the experiments confirm or disprove these theories.

An important note to conclude this Section. Apart from the systems described by the Real Dynamics or the Ideal Dynamics, there can exist systems that cannot be described in detail by either of them. The systems of this latter type correspond to Unpredictable dynamics, so that though their derailed description is theoretically possible, such description is experimentally unverifiable, and impossible to realize. Live system can possibly belong to this type.

## Section 5. Life and death.

We would like to note at the beginning of this Section, that it would be rather a set of hypotheses, whereas the previous ones were more or less strict.

We shall assume for the purposes of further discussion in this paper that Life completely conforms to the laws of physics. The following questions shall be discussed below:

What is Life and Death from the physical point of view?

Are there limitations on the possibility of describing live systems in terms of physics?

What distinguishes live systems from non-live ones from the physical point of view?

When do live systems develop consciousness and freedom of will from the physical point of view?

Life is usually defined as a highly organized form of existence of organic molecules having the capacity to perform certain functional activities, including metabolism, growth, reproduction, motion, adaptation, and responsiveness to the external stimulation. Live matter has the property of self-preservation over a long period of time, and can even increase its level of self-organization. This is a true, but too narrow definition: many of live systems have only part of these properties, and some of the properties can be found in non-live matter; besides, non-organic life forms are also possible.

Schroedinger was the first to attempt describing life from the point of view of modern physics [11]. He defined life as an aperiodic crystal, i.e. highly ordered form of matter (having therefore low entropy, and "feeding" not on energy but on negentropy of the environment, thus being an essentially open system). However, life is not based on simple repetition as the normal crystals are. Schroedinger further discussed two reasons making Real Dynamics of live systems stable against internal and external noise. These reasons are the statistical law of large numbers and the discreteness of quantum transitions, which ensures the stability of chemical bonds. Shroedinger stresses the similarity of the principles of live organisms' activity and clockwork operation: both create "order from order" despite the high temperature.

Soviet biophysicist E. Bauer noted that the high degree of order (and therefore low entropy) of live systems is expressed not only in the non-equilibrium distribution of matter in live organisms, but the structure of live matter itself is characterized by low entropy and high instability. This unstable structure is, on the one hand, maintained by the metabolism process, and on the other hand, catalyzes the metabolism. This assumption is only partially true, because, e.g., proteins and viruses preserve their structure even in the crystal form, but their low-entropy and highly unstable modifications inside the live organisms have this property. Nonetheless, the structure tends to degrade in time, thus making death inevitable and necessitating propagation to preserve life. Thus the metabolism process can only strongly delay the decay of the complex structure of live matter, and cannot keep it completely unchanged. The experimental results that Bauer quotes confirm energy release (and thus entropy increase), which results from autolysis, i.e. decay of live matter. Bauer suggests that the first stage of the autolysis process is caused by instability of the initial structure in the absence of the supporting metabolism process, and the second stage is caused by the action of protolytic (decomposing) ferments released or produced during autolysis. Bauer considered the surplus structural energy an essential characteristic of life.

Most of the above-mentioned papers considered only individual live organisms, whereas in fact it is possible to define and describe life as the totality of the live organisms (biosphere). Here the issue of the origins of life arises. The most complete answer to these questions from the point of view of modern physics is given in the paper of Elitzur [18]. The origin of life in this paper is considered as an ensemble of self-replicating molecules. Coming through the sieve of Darwinian natural selection, life accumulates the genetic information (or rather knowledge, in terms suggested by Elitzur) about the environment. This increases the level of the system's organization (negentropy) in agreement with the second law of thermodynamics. Lamarck's views in their too straightforward formulation were shown to contradict this law of physics. A wide range of issues is discussed in this paper. However, it has also the following drawbacks:

1) The description is true for life as a whole, i.e. as a phenomenon, but not so for an individual live organism.

2) The suggested reasoning disproves only the most simple and straightforward version of Lamarck's theory, whereas there is a number of hypotheses and experiments illustrating the possibility of realization of Lamarck's views in real life [32].
3) In Elitsur's view, the self-organizing dissipating systems suggested by Prigogine (e.g., Benard cells), in contrast to the live organisms, are denied the adaptation capability. Naturally, the adaptability of Benard cells is incomparable with that of live systems, but still they do exist, even if in a very primitive form. Thus, Benard cells change their geometry or even disappear depending on the temperature difference between the upper and lower layers of the fluid. This can be well considered a primitive form of adaptation.

An equilibrium ensemble in quantum mechanics in an equilibrium with the environment is described by a diagonal matrix in the energy representation.

Similarly, in classical mechanics there are no correlations between the molecules, which are the analogues of the non-diagonal elements of the density matrix.

Thus the disturbance of equilibrium has a twofold manifestation: first, non-equilibrium distribution of the diagonal elements and, second, non-zero non-diagonal elements of the density matrix (the latter corresponds to non-zero correlations in classical mechanics). Such correlations (or non-diagonal elements) are much more unstable, and decay much faster than the deviation of the diagonal elements from the equilibrium values.

Bauer determines life as a highly unstable system self-maintaining due to motion and metabolism. Therefore we can suppose that this instability is mostly due to the correlations (or non-diagonal elements in quantum mechanics), and that live systems tend to maintain these correlations and preserve them during time much greater than their natural relaxation time. In case of non-live system this can be achieves by simply isolating the system, whereas the open live systems actively interacting with the environment achieve this by external and internal motion and metabolism.

Note that live systems maintain correlations both between the internal elements, and between the system itself and the environment.

Further we introduce the concept of pseudo-alive systems. Simple physical systems displaying, in a primitive form, the real or supposed properties of the live systems would be called pseudo-alive systems. Thus, crystal growth models the ability of live systems to self-replication. By the way, the analysis of such systems allows finding a weak place in the Wigner's reasoning [6, 27] where he tries finding a controversy between the self-replicating capability and quantum mechanics.

Another example is quantum mechanical isolated systems that show the ability to preserve correlations, which is similar to maintaining high instability in live systems related to the preservation of correlations or non-diagonal elements of the density matrix.

However, this type of preservation is passive. Active retardation of relaxation of non-diagonal elements of density matrix, more similar to the way of maintaining correlations in live systems, can be observed in such open systems as micromasers [25]. Such systems are yet another example of pseudo-alive systems. Dissipative systems show the ability of preserving low entropy and primitive adaptation to the changing environment, similar to that of live systems.

The definition of life as the totality of systems maintaining correlation in contrast to the external noise, is a reasonable explanation of the mysterious silence of Cosmos, i.e. the absence of signals from other intelligent worlds. All the parts of the universe, having a unique center of origin (Big Bang), are correlated, and life maintains these correlations, which are at the base of its existence. Therefore the emergence of life in different parts of the Universe are correlated, so that all the civilizations have roughly the same level of development, and there are just no supercivilizations capable of somehow reaching Earth. The effects of long-range correlations can explain at least a part of the truly wonderful phenomena of human intuition and parapsychological effects. This explanation does not necessarily require the use of quantum mechanics, since similar correlations can be found in classical mechanics that has analogues of the non-diagonal elements of the density matrix. Attribution of such correlations exclusively to quantum mechanics is a common mistake.

Bohr made the next important contribution to understanding of life [26]. He stressed that the complete measurement of the system's state in quantum mechanics inevitably disturbs its behavior. This can be the reason of essential incognizability of life. Schroedinger's critic of these Bohr's views is not well-founded. This critic is based on the opinion that though a complete measurement of the system's state is possible in quantum mechanics, but it differs in its probabilistic nature from the measurement in classical case. The problem however does not consist in the impossibility of such a complete measurement. The essence of the problem is that such measurement changes further behavior of the system, which would have been different in the absence of the measurement [20]. The measurement changes fine correlations between the different parts of the system thus causing a change in its behavior. Note that this is also true for classical mechanics, where there is always a finite interaction between real systems.

Pseudo-alive systems illustrating how the measurement changes the system's dynamics are the oscillating almost isolated quantum systems that follow the pattern: A $\rightarrow$ (sum of A and B) $\rightarrow$ (difference of A and B) $\rightarrow$ A, where A and B are the pure states of the system.

Measuring the current state of the system (A or B) disrupts the sum or difference states thus changing the real dynamics of the system and deleting the correlations (non-diagonal elements of the density matrix) between the A and B that exist in the mixed states [10].

The advances in molecular genetics do not contradict this view. Developing the Real Dynamics of life is principally possible. In fact, real systems are essentially open systems actively interacting with their stochastic environment. The interaction of an external observer with such systems is usually much weaker, and cannot cause a major change in their behavior. However, attempting to achieve a too detailed understanding and description of life can disturb the fine and complex correlations. This can lead to the Non-predictable dynamics of live systems thus realizing the effect that Bohr predicted. It is possible that especially fine effects of human intuition and certain parapsychological effects belong to this unpredictability region. Since these effects can lie only in the narrow region bordering on the cognizability limits of the exact sciences, the corresponding capabilities cannot be amplified in the course of natural selection. This also limits our ability to reach a complete comprehension of these phenomena by means of science [21, 22].

## Section 6. Conclusions

Most of the real systems cannot be described with ideal equations of quantum or classical mechanics. The influence of the measurement and environment on the system (which is inevitable in quantum mechanics and almost always present in classical mechanics) disturbs the evolution of the system. The attempt to include the observer and the environment into the system to be described leads to the paradox of the self-observing system(see Appendix G). Such system cannot measure and retain complete information on its own state. Even the approximate (self-)description in such system can only have applicability in time domain limited by times small enough to be much less than the return time for this system. After this return time determined according to the Poincare's theorem all the information concerning the previous state of the system is inevitably erased. However, the description of the system in this case is possible in the frame of real dynamics. The possibility of such description is explained by the independence of real dynamics on the type and value of external noise for a wide range of noise types, thus real dynamics is determined only by the properties of the system itself. Real dynamics can be based on roughening of the distribution function or density matrix, because the initial state of the system is not defined precisely. The difference between the real and the ideal dynamics cannot be experimentally verified even when observer and environment are included into the system to be described, because self-description is limited in precision and observation time domain. Thus the returns of the system to the initial state predicted by the Poincare's theorem for ideal dynamics cannot be observed by the self-observing system due to the effect of erasing of information concerning the previous states. Introduction of real dynamics helps to solve all the known paradoxes of classical and quantum mechanics.

Thus this paper is not just an abstract philosophical discussion. Lack of understanding of the principles discussed in this paper can lead to mistakes in solving physical problems. Below several examples of such mistakes are given, including the errors in the pole theory for the problems of flame front motion and the "finger" growth at the liquid/liquid interface.

Sivashinsky *et al.* state that the Ideal Dynamics of poles causes an acceleration of the flame front propagation, and that this is not due to noise, because the effect does not disappear with the decrease of noise, and depends purely on the properties of the system.

However, the noise-connected Real Dynamics does not depend on the noise over a wide range of its values.

Tanveer *et al.* [24] found a discrepancy between the theoretical predictions for "finger" growth in the problem of interfacial fluid flow and the results of numerical experiments. However, no understanding was reached in paper [24] of the connection of this discrepancy with the numerical noise, which leads to a new Real Dynamics. These are just two examples taken from the daily practice of the author of the present paper, and more such examples can be easily found.

The results of this paper are necessary for through understanding of the basics of nonlinear dynamics, thermodynamics, and quantum mechanics.

**Appendix A. Density matrix**

Consider a beam of $N_a$ particles prepared in state $|\chi_a\rangle$, and another beam of $N_b$ particles in state $|\chi_b\rangle$ independent on the first one. To describe the combination of beams we introduce the mixed state operator ρ defined as follows:

$$\rho=W_a|\chi_a\rangle\langle\chi_a| + W_b|\chi_b\rangle\langle\chi_b|,$$

where $W_a=N_a/N$, $W_b=N_b/N$, $N=N_a+N_b$

Operator ρ is called density operator or statistical operator. It describes the way the beams were prepared and therefore contains a complete information about the total beam. In this sense the mixture is completely defined by the density matrix. In the special case of a pure state $|\chi\rangle$ the density operator is given by the expression

$\rho=|\chi\rangle\langle\chi|$.

Operator ρ is usually convenient to write in matrix form. Therefore we choose a basic set of states (most commonly used are $|+1/2\rangle$ and $|-1/2\rangle$) and decompose the $|\chi_a\rangle$ and $|\chi_b\rangle$ states over this basic set as follows:

$$|\chi_a\rangle=a_1^{(a)}\,|+1/2\rangle+a_2^{(a)}\,|-1/2\rangle,$$

$|\chi_b\rangle = a_1^{(b)}\,|+1/2\rangle + a_2^{(b)}\,|-1/2\rangle$.

In the representation of $|\pm 1/2\rangle$ states we have the relations for the ket states:

$$|\chi_a\rangle = \begin{pmatrix} a_1^{(a)} \\ a_2^{(a)} \end{pmatrix}$$

$$|\chi_b\rangle = \begin{pmatrix} a_1^{(b)} \\ a_2^{(b)} \end{pmatrix},$$

and for the conjugated states:

$$\langle\chi_a| = (a_1^{(a)*}\,,\ a_2^{(a)*}\,),$$
$$\langle\chi_b| = (a_1^{(b)*}\,,\ a_2^{(b)*}\,).$$

Using the matrix multiplication rules we obtain for the "external product":

$$|\chi_a\rangle\langle\chi_a| = \begin{pmatrix} a_1^{(a)} \\ a_2^{(a)} \end{pmatrix} (a_1^{(a)*}\,,\ a_2^{(a)*}\,) = \begin{pmatrix} |\,a_1^{(a)}\,|^2 & a_1^{(a)} a_2^{(a)*} \\ a_1^{(a)*} a_2^{(a)} & |\,a_2^{(a)}\,|^2 \end{pmatrix}$$

and a similar expression for the $|\chi_b\rangle\langle\chi_b|$ product. Substituting these expressions into the density operator, we obtain the density matrix.

$$\rho = \begin{pmatrix} W_a\left|a_1^{(a)}\right|^2 + W_b\left|a_1^{(b)}\right|^2 & W_a a_1^{(a)} a_2^{(a)*} + W_b a_1^{(b)} a_2^{(b)*} \\ W_a a_1^{(a)*} a_2^{(a)} + W_b a_1^{(b)*} a_2^{(b)} & W_a\left|a_2^{(a)}\right|^2 + W_b\left|a_2^{(b)}\right|^2 \end{pmatrix}$$

Since the $|\pm 1/2\rangle$ states were used for the basic state, the obtained expression is called density matrix in $\{|\pm 1/2\rangle\}$ representation.

In conclusion, we make several notes concerning the statistical matrix $P_0$, which has remarkable properties. We know that all the possible macroscopic states of the system in classical statistical thermodynamics are considered *a priori* equiprobable. In other words, the states are considered equally probable, unless information is available concerning the total energy of the system, its contact with the thermostat ensuring the constant temperature of the system, etc. Similarly, in wave mechanics all the states of the system corresponding to the functions forming the complete system of orthonormalized functions can be *a priori* considered equiprobable. Let $\varphi_1,\ldots,\varphi_k$, is such a system of basis functions. Provided the system is characterized by a mixture of the $\varphi_k$ states, in the absence of other relevant information we can assume that the statistical matrix of the system has the form

$$P_0 = \sum_k p P_{\varphi_k}\ , \text{ where } \sum_k p = 1\,,$$

i.e. that $P_0$ is the statistical matrix of a mixed state with all equal statistical weights. Since $\varphi_k$ are the basis functions, matrix $P_0$ can be represented as follows:

$$(P_0)_{kl} = p\delta_{kl}$$

If matrix $P_0$ characterizes the statistical state of the ensemble of systems at the initial moment of time, and the same value A is measured in all the systems of the ensemble, then the statistical state of the ensemble would be still characterized by the $P_0$ matrix.

The equations of motion for the density matrix ρ have the form:

$$i\frac{\partial \rho_N}{\partial t} = L\rho_N ,$$

where *L* is the linear operator:

$L\rho=H\rho-\rho H=[H,\rho]$,

where *H* is the energy operator of the system.

If A is the operator of a certain observable, then the average value of the observable can be found as follows:

$\langle A\rangle = \mathrm{tr} A\rho$

**Appendix B. Reduction of the density matrix and the theory of measurement**

Assume that the states $\sigma^{(1)}$, $\sigma^{(2)}$, ... are "clearly discernible" in measuring a certain object. The measurement performed over the object in one of these states yields numbers $\lambda_1$, $\lambda_2$, ... . The initial state of the measuring device is designated *a*. If the measured systems was initially in the state $\sigma^{(v)}$, then the state of the complete system "measured system plus the measuring device" before their interaction is determined by the direct product a X $\sigma^{(v)}$. After the measurement

$$a \times \sigma^{(v)} \rightarrow a^{(v)} \times \sigma^{(v)}$$

Assume now that initial state of the measured system is not clearly discernible, but is an arbitrary mixture: $\alpha_1 \sigma^{(1)} + \alpha_2 \sigma^{(2)}+$... of such states. In this case, due to the linearity of the quantum equations, we obtain:

$$a \times [\Sigma\alpha_v\sigma^{(v)}] \rightarrow \Sigma\alpha_v[a^{(v)} \times \sigma^{(v)}]$$

There is a statistical correlation between the state of the object and the state of the device in the final state resulting from the measurement. A simultaneous measurement of two values in the system "measured object and measuring system" (the first one is the measured characteristic of the studied object, and the second is the position of the measuring device indicator) always leads to correlating results. Therefore one of the measurements mentioned above is superfluous: a conclusion on the state of the measured object can always be made based on observing the measuring device.

The state vector obtained as the measurement result cannot be represented as a sum in the right hand part of the relation above. It is a so-called mixture, i.e. one of the state vectors having the form:

$$a^{(v)} \times \sigma^{(v)},$$

and the probability of this state appearing as the result of interaction between the measured object and the measuring device is $|\alpha_v|^2$. This transition is called wave packet reduction. And corresponds to the transition of the density matrix from the non-diagonal form $\alpha_v\alpha_\mu^*$ to the diagonal form $|\alpha_v|^2 \delta_{v\mu}$. This transition is not described by the quantum mechanical equations of motion.

**Appendix C. Phase density function**

The state of a system of N particles identical from the point of view of classical mechanics can be given by the coordinates $\mathbf{r}_1, \ldots, \mathbf{r}_N$ and momenta $\mathbf{p}_1, \ldots, \mathbf{p}_N$ of all the N particles of the system. For brevity's sake we shall further use the notation

$x_i=(\mathbf{r}_i,\mathbf{p}_i)$ $(i=1, 2, \ldots, N)$

to designate the set of coordinates and the momentum spatial components of a single particles, and the designation

$X= (x_1, \ldots, x_N)=(\mathbf{r}_1, \ldots, \mathbf{r}_N, \mathbf{p}_1, \ldots, \mathbf{p}_N)$

to denote the set of coordinates and momenta of all the particles of the system. The corresponding state of 6N variables is called 6N-dimensional phase space.

We shall consider a Gibbs ensemble, i.e. a set of identical macroscopic systems in order to define the concept of distribution function. The experimental conditions are similar for all these systems. However, since these conditions do not determine the state of the systems unambiguously, then different values of X shall correspond to different states of the ensemble at a given time *t*.

We select a volume dX in the vicinity of the point X. Assume that at a given time *t* this volume contains points characterizing the states of dM systems from the total number M of systems in the ensemble. Then the limit of the ratio of these values

$$\lim_{m\to\infty} dM/M=f_N(X,t)dX$$

defines the density function of the distribution in phase state at time *t*. This function is obviously normalized as follows:

$$\int f_N(X,t)dX=1$$

The Liouville equation for the phase density function can be written in the form:

$$i\frac{\partial f_N}{\partial t} = Lf_N = \{H, f_N\}$$

where *L* is the following linear operator:

$$L = -i\frac{\partial H}{\partial p}\frac{\partial}{\partial x} + i\frac{\partial H}{\partial x}\frac{\partial}{\partial p},$$

where *H* is the energy of the system.

**Appendix D. Coarsening of the phase density function and the molecular chaos hypothesis.**

Coarsening of the density function is called its substitution with an approximate value, e.g.:

$$f_N^*(X,t)=\int_{(y)} g(X-Y)f_N(Y,t)$$

where

$$g(X)=1/\Delta\ D(X/\Delta)$$

$D(x)= 1$ for $|X|<1$
$D(x)= 0$ for $|X|\geq 1$

Another example of coarsening is the “molecular chaos hypothesis”. It implies substitution of a two-particle distribution function with a product of single-particle functions as follows:

$f(x_1, x_2, t) \to f(x_1, t) f(x_2, t)$

**Appendix E. Definitions of entropy.**

We can give the following definition of entropy:

$S = -k \int_{(X)} f_N(X,t) \ln f_N(X,t)$

In quantum mechanics entropy is defined via density matrix:

$S = -k \, tr \, \rho \ln \rho$ [29]

where *tr* stands for matrix trace.

Entropy thus defined does not change in the course of reversible evolution. Coarsened values of $f_N$ or ρ should be used to obtain the changing entropy.

**Appendix F. Prigogine’s New Dynamics.**

The New dynamics introduced by Prigogine is often mentioned in the present paper. Below a brief introduction to this theory is given based on the monographs [9, 12].

A linear operator Λ is introduced, which acts on the phase density function or density matrix ρ so that:

$$\tilde{\rho} = \Lambda^{-1}\rho$$
$$\Lambda^{-1} 1 = 1$$
$$\int \tilde{\rho} = \int \rho$$

where $\Lambda^{-1}$ remains positive. The requirement on the operator Λ is that the function Ω defined via the function $\tilde{\rho}$ as follows:

$\Omega = tr\tilde{\rho}^{+}\tilde{\rho}$ or $\Omega = tr\tilde{\rho} \ln\tilde{\rho}$

complies with the inequality $d\Omega/dt \leq 0$

The equation of motion for the transformed function $\tilde{\rho}$ is

$$\frac{\partial\tilde{\rho}}{\partial t} = \Phi\tilde{\rho}$$

where $\Phi = \Lambda^{-1} L \Lambda$

Φ is noninvertible markovian semigroup.

$\Lambda^{-1}(L) = \Lambda^{+}(-L)$

Operator $\Lambda^{-1}$ for the phase density function corresponds to the coarsening in the direction of phase volume decrease. In quantum mechanics such operator can only be found for an infinite volume or an infinite number of particles. A projection operator P is introduced in quantum mechanics, which makes all the non-diagonal elements of the density matrix zero. Operator Φ and the basis vectors of the density matrix are chosen so that the operators Φ and P are permutable:

$$\frac{\partial \mathrm{P}\tilde{\rho}}{\partial t} = \mathrm{P}\Phi\tilde{\rho} = \Phi\mathrm{P}\tilde{\rho}$$

**Appendix G. Impossibility of self-prediction of system’s evolution.**

Suppose there is a powerful computer able to predict its own future and that of its environment based on the calculation of motion of all the molecules. Suppose that the prediction is rolling of a black or white ball from a certain device, which is an integral part of the computer and is described by the machine. The device rolls out a white ball when the computer predicts a black one, and rolls out a black ball when the white one is predicted. It is clear that the predictions of the computer are always false. Since the choice of the environment is arbitrary, then this contrary instance proves the impossibility of exact self-observation and self-calculation. Since the device always contradicts the predictions of the computer, then complete self-prediction of the system including both the computer and the device is impossible.

**Appendix H. Table of correspondence between the quantum and classical mechanics.**

| **Quantum mechanics** | **Classical mechanics** |
|---|---|
| Density matrix | Phase density function |
| Equation of motion for the density matrix | Liouville equation |
| Wave packet reduction | Coarsening of the phase density function or the molecular chaos hypothesis |
| Unavoidable interaction of the measured system with observer or environment, described by reduction | Theoretically infinitesimal but in reality a finite, if small, interaction of the measured system with observer or environment |
| Non-zero non-diagonal elements of the density matrix | Correlations between the velocities and positions of particles in different parts of the system |

# Основы нелинейной динамики или Реальная Динамика, Идеальная Динамика, Непредсказуемая Динамика и “Шредингеровский кот”.

Купервассер Олег
Технион, Хайфа, Израиль

**Абстракт.**

В статье обсуждены парадоксы лежащие в основе термодинамики и квантовой механики. Дано их разрешение с точки зрения влияния внешнего наблюдателя (окружающей среды), разрушающего корреляции системы, или ограниченности самопознания системы в случае, когда и наблюдатель, и окружающая среда включены в систему. Введены понятия Реальная Динамика, Идеальная Динамика и Непредсказуемая Динамика. Дано рассмотрение явления жизни с точки зрения этих Динамик.

## Введение.

Статистическая механика и квантовая механика - две разработанные и хорошо известные теории, основа современной физики. Тем не менее они содержат ряд парадоксов, которые заставляют многих ученых усомнится во внутренней непротиворечивости этих теорий. Перечислим их:

1) Базисные уравнения классической и квантовой механики обратимы во времени в то время как законы термодинамики, например, закон возрастания энтропии в замкнутых системах, необратим, хотя он, по сути дела, должен выводиться из этих уравнений [1], [2].
2) Тоже относится к противоречию между законом возрастания энтропии (Приложение E) в замкнутых системах и законом Пуанкаре о возвратах замкнутой системы сколь угодно близко к начальному состоянию [1], [2].
3) Парадокс "Шредингеровского  кота" , т.е. редукция волнового пакета (Приложение B) и переход из чистого в смешанное состояние в макроскопических системах или в процессе измерения квантовой системы. Этот процесс не описывается уравнениями квантовой механики, совершенно непонятно в какие моменты времени он происходит и какова его длительность.Тем не менее, без него квантовая механика неполна [3].

4) Микроскопическая энтропия (Приложение E) классических систем,определенная через функцию плотности в фазовом пространстве, остается неизменной.(Аналогично энтропии квантовых систем, определенной через матрицу плотности [15]) Макроскопическая энтропия (Приложение E) систем, определенная через макроскопические переменные или, что математически эквивалентно, «огрубленную» (Приложение D)  для классической механики функцию плотности (Приложение C)  в фазовом пространстве ( и редуцированную  матрицу плотности (Приложение A)  для квантовой механики), может как возрастать, так и убывать с одинаковой вероятностью из-за обратимости движения.(В отличие от иногда высказываемого заблуждения, что процессы с убыванием энтропии менее вероятны. На самом деле, наиболее вероятны состояния с неизменной и максимальной для данной системы энтропией, тогда как состояния, ведущие как к убыванию, так и возрастанию энтропии одинаково редки по сравнению с этими равновесными состояниями) Почему, в таком случае, в реальном мире во всех  макроскопических классических системах всегда наблюдается лишь возрастание энтропии? Почему, чтобы получить законы для  этого возрастания энтропии нам приходится вводить некие дополнительные предположения ( вроде, например, гипотезы молекулярного хаоса Больцмана (Приложение D)  или эквивалентного ему по смыслу огрубления функции (Приложение D)   плотности в фазовом пространстве), помимо основных уравнений классической механики, хотя они и дают нам ее полное описание? [1], [2]

5)Для описания макроскопического состояния используются те или иные макроскопические переменные. Теоретически их выбор ничем не ограничен, это может быть любая функция микроскопических параметров, тем не менее, как правило, выбор "удобных" переменных
достаточно предопределен. Чем?

6)Существует множество живых систем, в том числе обладающих сознанием и свободой воли. Какие физические свойства принципиально отличают живые (сознающие) системы от неживых? Насколько полно они могут быть описаны известными уравнениями физики и термодинамики? Не даром Шредингер использовал для формулировки своего парадокса в качестве макроскопического детектора именно живого кота! Некая связь между описанными выше базисными парадоксами физики и тайнами жизни интуитивно ясна многим физикам.Стоит попытаться сформулировать  ее более ясно.

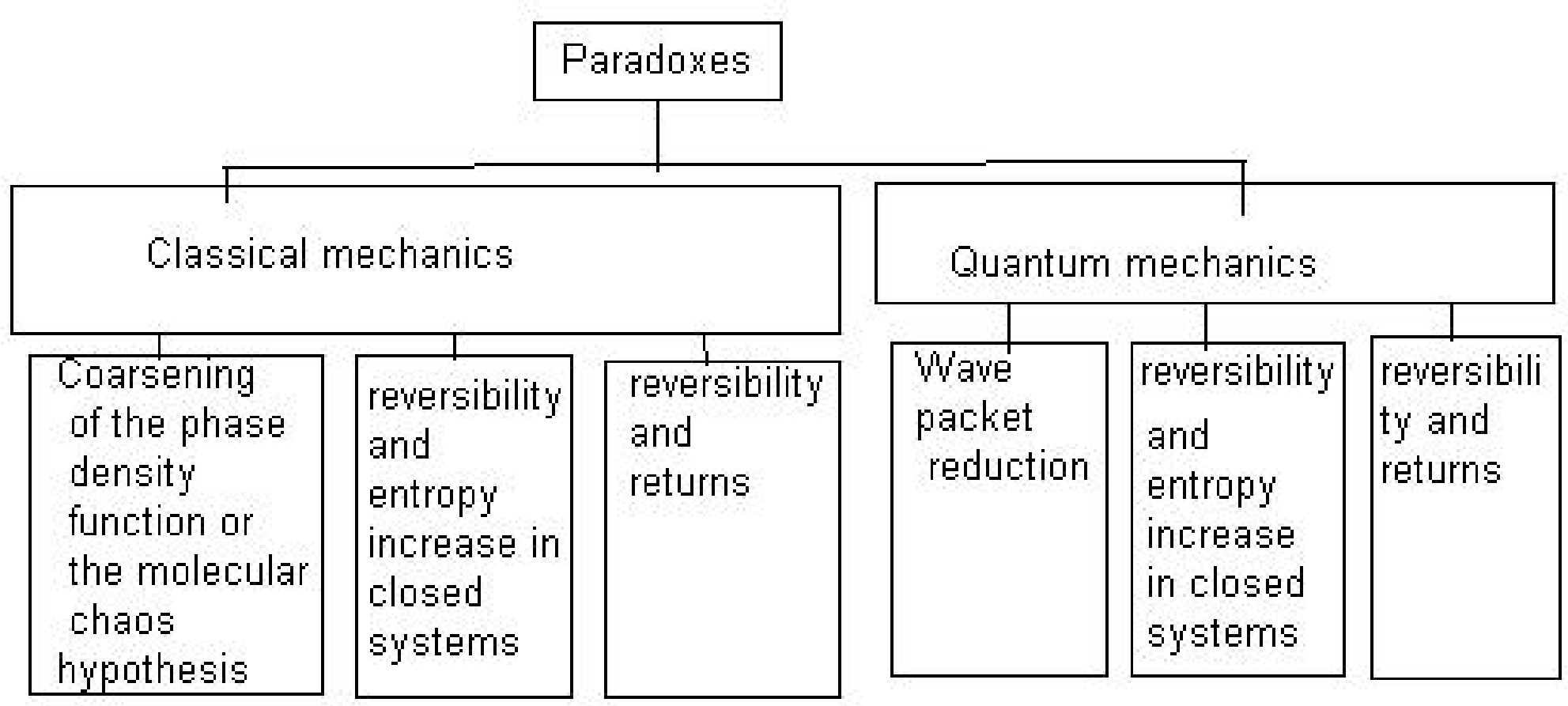

Figure1 Paradoxes in Classical and Quantum mechanics

Разрешение этих парадоксов не требует изобретения новых законов физики, хотя их экспериментальная проверка и может привести к ним.
Мы будем анализировать эти проблемы строго в рамках известных законов физики.

Имеется обширная литература посвященная этим парадоксам [1], [2], [3]. Многое из сделанного верно. Проблема в том, что эти рассмотрения не дают ключевого свойства динамической системы, приводящего ко всем этим парадоксам и дают лишь фрагменты их полного решения.

Ситуация тут напоминает парадокс Гиббса [4], где скачок энтропии при смеси газов объясняли отсутствием переходных форм между одинаковыми и разными газами. Тем не менее такие переходные формы возможны, величину скачка можно регулировать и рассмотрение ( не поверхностное) дает не только глубокое понимание и разрешение этого парадокса, но и более глубокое понимание самих основ физики и не требует введения новых законов.

Аналогичная ситуация складывалась и со спином электрона: учебники переполнены утверждениями, что его невозможно интерпретировать как собственный момент именно физического (т.е. в пространстве) вращения частицы. Забавно, тем не менее, что в рамках уравнения Дирака это прекрасно делается и спин может быть истолкован как собственный момент чисто физического вращения волновой функции [5].

Также и вышеуказанные парадоксы могут быть при аккуратном рассмотрении глубоко поняты и дать богатый материал для понимания основ существующей физики с присущими ей ограничениями для объяснения основ окружающего мира. Физика не столь всесильна даже чисто теоретически (как почти все, даже профессиональные физики, уверены) и эти ее ограничения вполне можно вполне вывести из нее самой, не изобретая новых теорий. Эти ограничения и ведут к указанным выше парадоксам. Что интересно, при всем их разнообразии и внешней несвязанности все они имеют единый корень и причину!

Инструментом для нашего рассмотрения будет Нелинейная Динамика, наука которая ставит себе целью НЕ изобретение новых фундаментальных законов физики, а используется для более глубокого понимания уже существующих и для нахождения общих свойств и законов   относящихся к совершенно казалось бы разным физическим системам (в смысле предметной области их приложения), но, тем не менее, имеющим очень похожую динамику.

**Часть 1. Квантовый парадокс «Шредингеровского кота».**

Не  будем здесь подробно описывать условия этого парадокса, они хорошо описаны во многих источниках [14], [3], [6], [7]. Остановимся здесь на его сути. Кот - это макроскопическая система, которая, будучи описана микроскопически (квантово), может находится одновременно в двух состояниях - кот живой и кот мертвый, являясь их «взвешенной суммой» (т.н. волновым пакетом пси-функций)! В реалии, как возможный сторонний наблюдатель, так и сам кот своим сознанием фиксируют всегда лишь одно из этих двух состояний с вероятностью, определяемой квадратами их «весов» в вышеописанной «взвешенной сумме». На языке математики это соответствует переходу от чистого к смешанному состоянию и диагонализации матрицы плотности (Приложение A,B)  .
Этот процесс носит название редукции волнового пакета и не описывается уравнениями квантовой механики. Более того, он соответствует возрастанию микроскопической энтропии (определенной с помощью матрицы плотности), увязываясь тем самым с предыдущими двумя парадоксами и вступая в противоречие с обратимостью квантовых уравнений и "возвратами" квантовой системы. Поскольку эти парадоксы присущи и классической механике
это позволит нам найти в дальнейшем также и классический аналог для парадокса «шредингеровского кота», который прекрасно существует, несмотря на всеобщую полную убежденность в его "чисто квантовой природе", связанную с отсутствием глубокого понимания сути и причин этих парадоксов. Необходимость введения этой загадочной редукции волнового пакета для макроскопических систем (являющимися, как правило, и конечными объектами в измерительных системах, т.е. их детекторами) и является основой и сутью парадокса «шредингеровского кота».

Кстати, в литературе встречаются попытки описать средствами
искусства некие состояния сознания, отвечающие квантовой «взвешенной сумме» [8]. Но если таковые и существуют на самом деле, они являются скорее экзотикой, чем правилом.

Для разрешения парадокса, казалось бы, проще всего было бы предположить, что эволюция просто распадается на две возможные «ветви эволюции» с определенной вероятностью каждая, аналогично тому, как это происходит в статистической механике, где состояние системы описывается не точкой в фазовом пространстве, а является неким «облаком» таких точек. Дело в том, что редукция волнового пакета, в отличие от «разделения на ветви» эволюции в статистической механике, приводит к иной эволюции ветвей, чем она была бы, сохрани мы «взвешенную сумму». Т.е. эти две возможные ветви эволюции оказывают взаимное влияние друг на друга через смешение результатов итогового измерения (которое по сути и является редукцией), проводимого для «взвешенной суммы» этих ветвей, поскольку решение является арифметической суммой, а не статистической смесью этих ветвей.

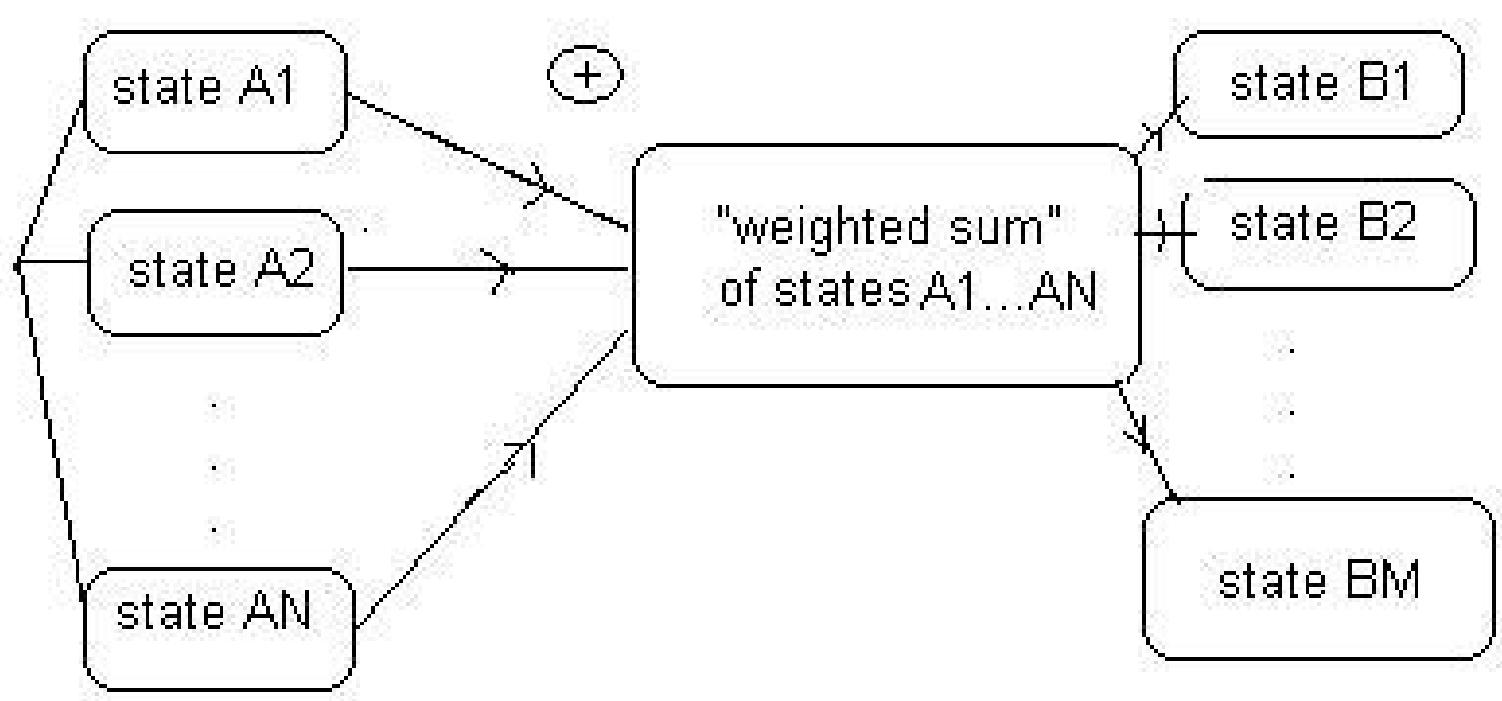

Fugure 2a
"Weighted sum" evolution of states correspondent to different values of macroscopic variable A after measurement of macroscopic variable B and "wave packet" reduction.

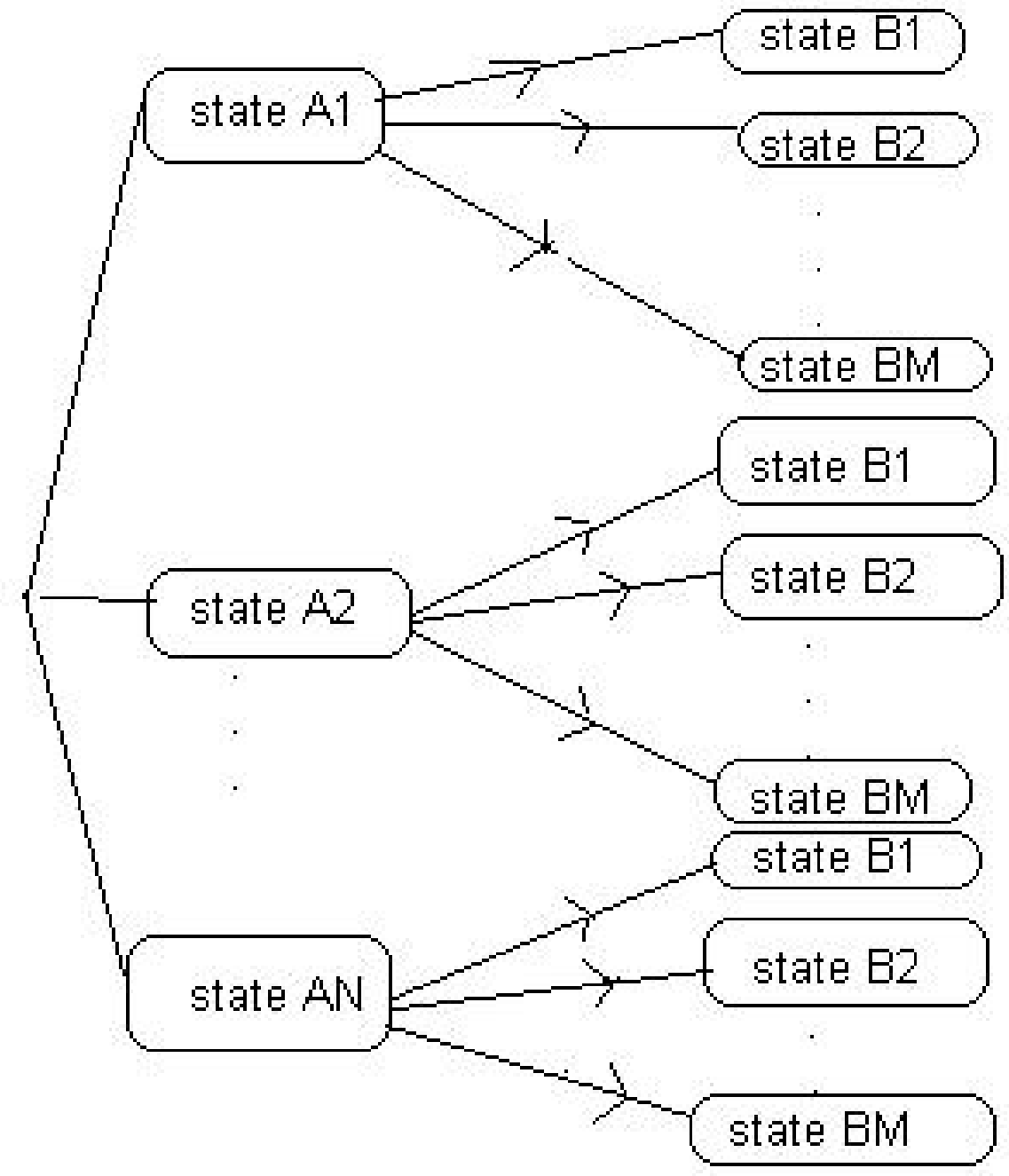

Figure 2b Solution is a statistical sum of the branches, and not "weighted sum" reduction as on Figure 2a.

Это находит подтверждение в следующих примерах. Первый пример связан с обратимостью квантовых уравнений движения. Как уже было сказано, редукция волнового пакета приводит к возрастанию энтропии, в то время как обратимая квантовая эволюция оставляет ее неизменной. Вряд ли можно рассчитывать на то, что эти два, столь разные, типа эволюции дадут одинаковый результат.

Второй пример связан с теоремой Пуанкаре о «возвратах» динамической системы [1], [2]. Хотелось бы здесь остановиться на некоторых особенностях этой теоремы для квантовых систем. В случае классической механики большинство динамических систем систем являются хаотическими: времена возврата образуют случайную последовательность и сильно меняются по величине при малейших изменениях начальных условий. Тем не менее существует хотя и малый, но физически значимый класс систем с периодическими или почти периодическими временами возврата, устойчивых к малым погрешностям начальных условий. Эти системы интегрируемы в переменных действие-угол.

В случае квантовых систем ситуация прямо обратная - замкнутая система с ограниченным объемом и с конечным числом частиц - всегда обладает периодическими или почти периодическими временами возврата, кроме систем бесконечного объема или бесконечного числа частиц, где это время возврата равно бесконечности [9].

Соответственно, возвращаясь к нашим уравнениям, редукция кота приводит к возрастанию микроскопической энтропии, делая невозможным возврат, поскольку он соответствует прежней, меньшей энтропии, а обычная обратимая квантовая эволюция оставляет энтропию неизменной. Таким образом, редуцированная и не редуцированная системы имеют различную динамику.

Третий забавный пример - «котелок, который никогда не закипит» [3]. Пусть мы имеем частицу, которая должна перейти с верхнего на нижний уровень энергии в соответствие с законами квантовой механики (например распасться). Если проводить акты наблюдения за ней слишком часто она никогда этого не сделает! Искажение эволюции системы редукцией, происходящей в моменты наблюдения, приводит к этому эффекту. Как часто происходят эти эти акты редукции в реальности и какова их собственная временная продолжительность? Не ясно.
Кстати, этот парадокс объясняет, почему распад частиц или состояний в квантовой механике не является точно экспоненциальным, а лишь близким к нему. Эти отклонения, казалось бы, позволяют определить «возраст» системы, поскольку лишь экспоненциальный тип распада делает это невозможным. Тем не менее это не так - сам процесс наблюдения и вносимые им искажения из-за редукции делают разницу между экспоненциальным и квантовым типом распада ненаблюдаемой.

Эти рассуждения доказывают, что редукция - это не просто механическое разделение движения на две возможные независимые ветви, она меняет всю реальную динамику системы.

Рассмотрим возможные известные подходы к решению парадокса кота.Все эти подходы представляют собой некие правильные шаги в нужном направлении, но не приводят к полному разрешению парадокса.

Последнее время появилось огромное количество различных интерпретаций квантовой механики, самая известная и популярная из которых «многомировая» [3], которая предполагает, что в момент редукции волнового пакета не выбирается лишь одна ветвь, а они существуют одновременно в неких параллельных мирах, в каждом из этих миров наблюдатель видит лишь одну из ветвей, тем не менее, поскольку они существуют одновременно, когда проводится измерение, все возможные «миры» оказывают влияние на результаты этого измерения, обеспечивая взаимное влияние ветвей, вытекающее из законов квантовой механики, которое при стандартной редукции уничтожается выбором лишь одной из ветвей. Проблема этой интерпретации (впрочем как и всех иных) в том, что она не разрешает сами трудности, а лишь переносит их в иную, менее заметную для нас, плоскость. Так, наблюдатель в любом из возможных «миров» имеет полную информацию лишь о нем самом и ничего в явной форме не знает о других мирах, что делает фактически его информацию неполной и не достаточной для предсказания результатов измерения, которые определяются итоговой эволюцией во всех «мирах». Таким образом, нарушение нормальной квантовой эволюции в случае редукции заменяется принципиальной непредсказуемостью результатов измерения из-за ограниченной информации в каждом из «миров». Хотя намек на правильное разрешение парадокса кота тут есть.

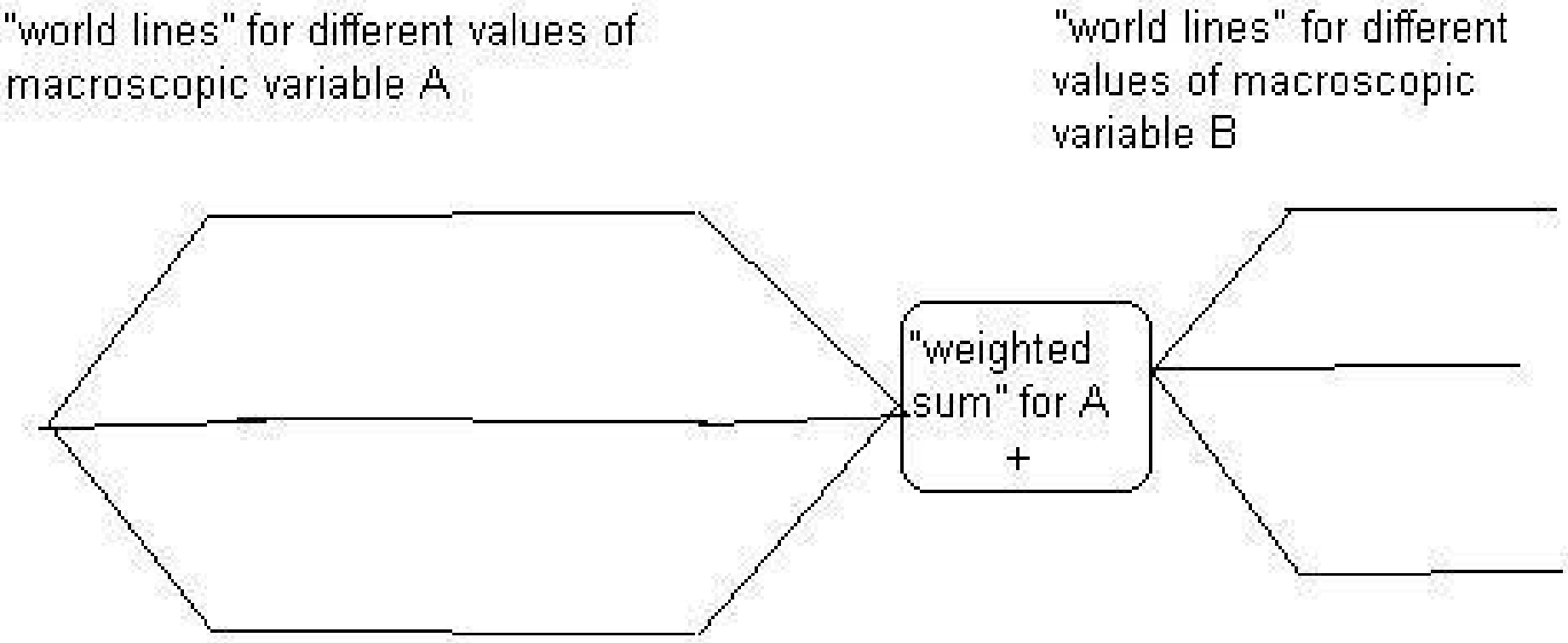

Figure 3 Many-world concept.Possible evolution branches exist in simultaneously parallel "worlds", and the reduction does not call for choosing a single branch, because the observer can see only one branch in each "world".
However, since all the branches exist at the same time, all the possible "worlds" influence the result of the measurement and thus ensure the interaction of the branches. So the problem of such many-world concept is that the observer in any "world" have information only about this "world" and can't predict the result of the measurement, that depends on all "worlds".

Следующим подходом к разрешению парадокса кота является учет влияния внешнего наблюдателя. Ведь редукция волнового пакета и происходит в момент наблюдения или измерения.Действительно, редукция и нарушение нормального хода квантовой эволюции может быть объяснено в этом случае незамкнутостью системы и воздействием внешних сил.

Тут нам следует остановится на одном из главных принципиальных моменте, отличающем классические и квантовые системы. В классических системах влияние измерительного прибора на исследуемый процесс может быть (конечно чисто теоретически, а не практически) сделано сколь угодно малым. В квантовой механике, измерение не может быть проведено без редукции волновой функции и хоть и малого, но вполне конечного воздействия на измеряемую систему.Чем точнее мы хотим измерить состояние квантовой системы, тем более сильное возмущение мы в нее вносим. Это делает квантовую механику принципиально не замкнутой.Именно в этом, а не в ином типе базисных уравнений или вероятностном характере ее законов, как принято обычно считать, и состоит принципиальное отличие квантовой и классической механик. Следует особо отметить, что это отнюдь не делает невозможным точное обратимое квантовое описание систем или их экспериментальную проверку внешним наблюдателем.Просто это возможно в ограниченной мере, лишь при специфических условиях, которых мы коснемся ниже. Здесь лишь отметим, что возможна проверка внешним наблюдателем лишь правильности самих обратимых квантовых уравнений движения, но не само точное изменение параметров и истории эволюции системы, ими описываемых, поскольку они определяются не только видом уравнений, но и начальным состоянием системы, неизбежно нарушаемым наблюдением.

Невозможность исключить влияние измерения иллюстрируется опытом с квантом, падающим на экран с двумя щелями, и интерференционной картиной за ними: при измерении через какую из двух щелей прошел квант влияние второй щели становится невозможным исходя из максимальности скорости света и принципа причинности, и интерференционная картина исчезает, заменяясь простой суммой интенсивностей сигналов от двух щелей.

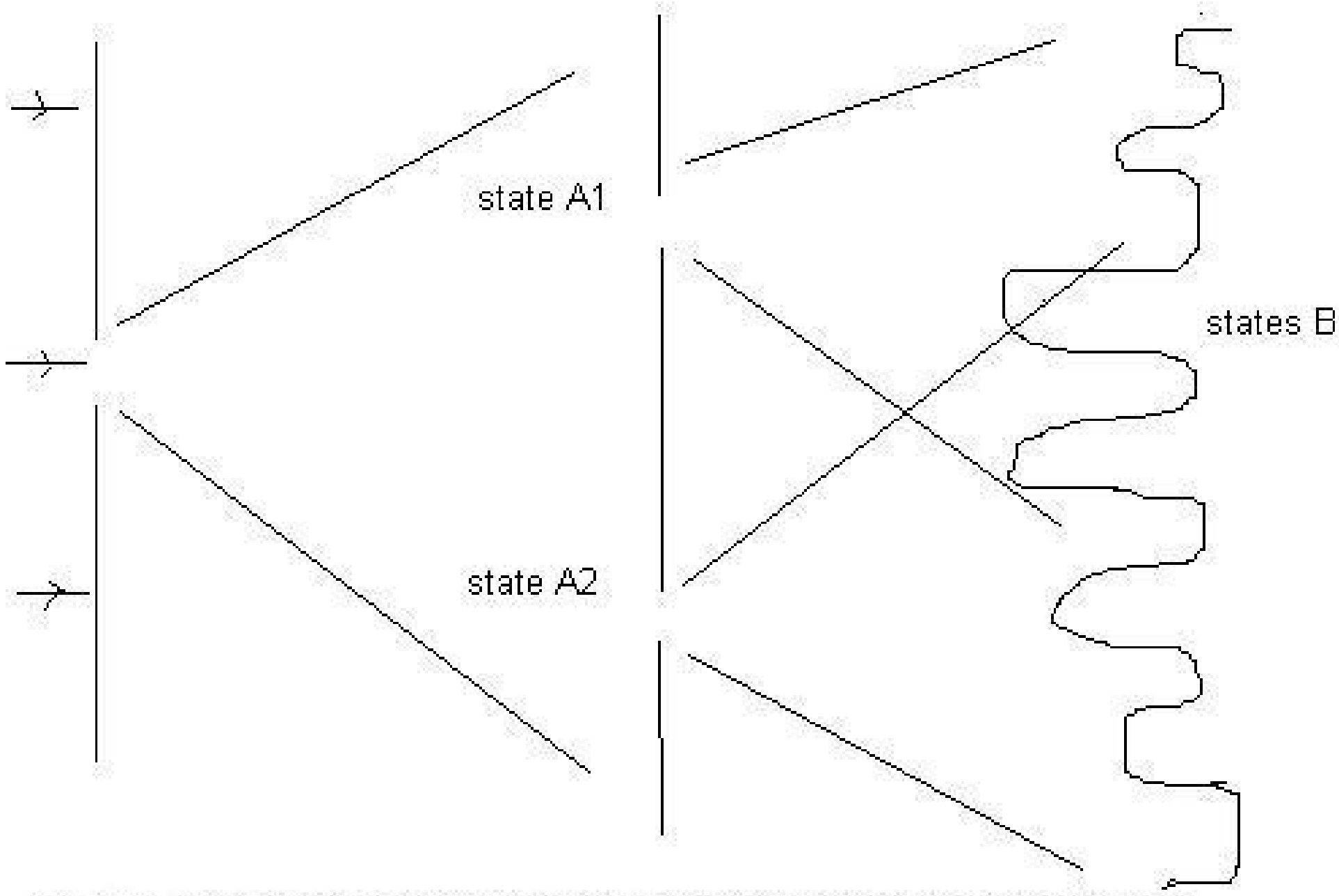

Fifure4a interference pattern from the two slits in the case of no detection, , which of the two slits the quantum passes through
example of the case on Fugure 2a

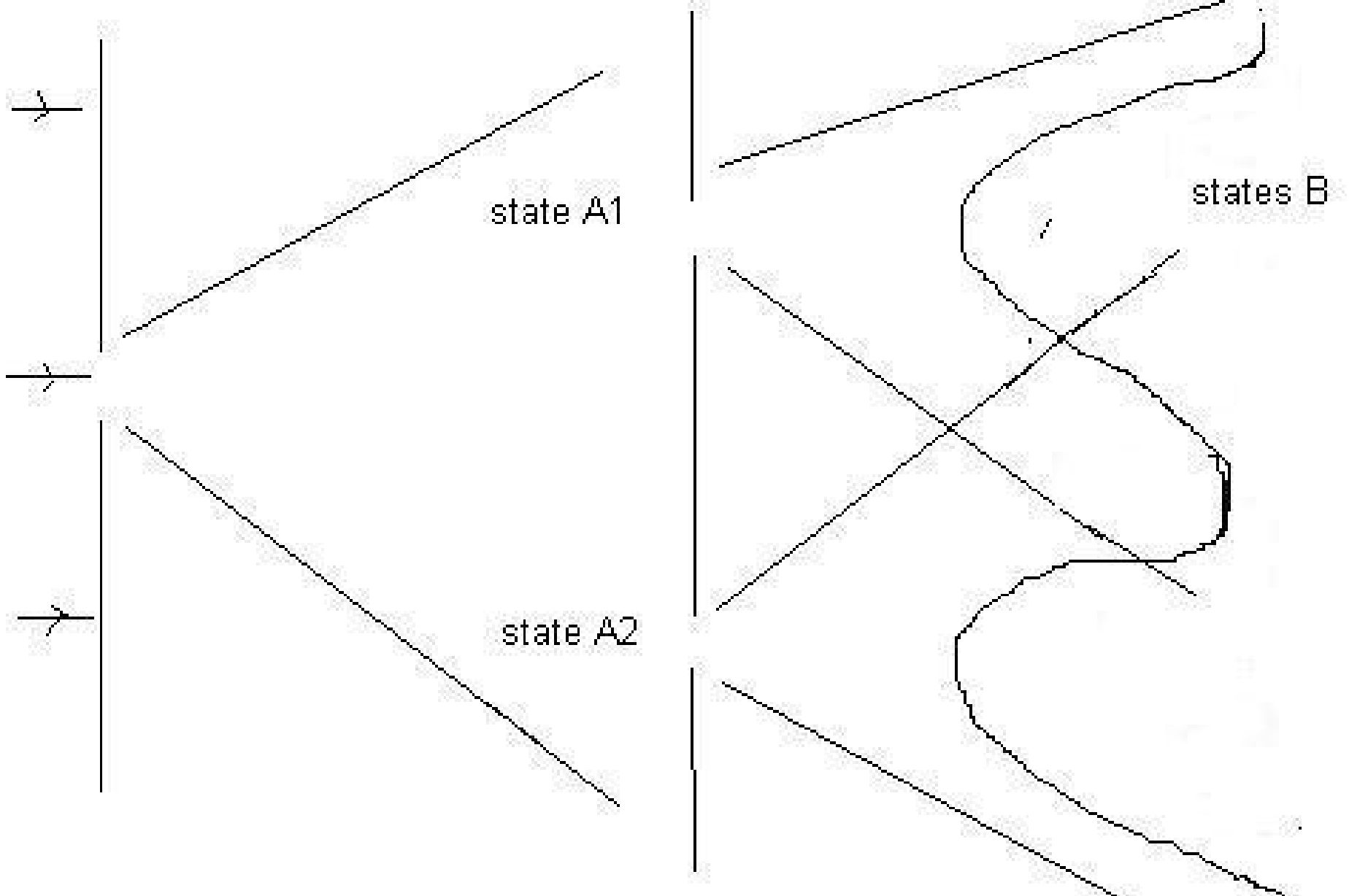

Figure 4b The sum of intensities of the signals from the two slits in the case of detection, which of the two slits the quantum passes through
Example of the case on Figure 2b

Для придания процессу взаимодействия с наблюдателем объективности и независимости от субъективного наблюдателя его воздействие может быть заменено неконтролируемым взаимодействием с макроскопической окружающей средой, что также вносит элемент незамкнутости в систему. Этот процесс носит название декогеренизации [16] и пользуется заслуженной популярностью. Его главное преимущество состоит в том, что он заменяет набор квантовых обратимых уравнений движения для замкнутой системы и таинственный необратимый процесс редукции замкнутой системой уравнений, включающей внешний шум.

Эти уравнения имеют конкретное и чисто практическое значение. Здесь-то нас и поджидают новые трудности. Менее принципиальная состоит в том, что параметры системы и уравнения движения становятся зависимыми от конкретного вида некого неконтролируемого внешнего воздействия. Каким оно должно быть выбрано, чтобы обеспечить верное и правильное описание системы?

Гораздо более принципиальная трудность состоит в том, что как наблюдатель, так и внешняя среда могут быть в принципе включены в саму систему и процесс редукции, который реально наблюдается и никуда не исчезает и в этом случае, будет происходить уже в замкнутой системе.В этом случае, найденное выше объяснение редукции, как некого внешнего воздействия, отпадает и мы возвращаемся к нашему парадоксу в его прежней нерешенной форме. Простейшим примером для понимания этого явления является случай, когда сам кот занимается самонаблюдением, констатируя, что он сам еще жив, а не мертв.В этом случае конечным этапом измерения будет сознание кота. Что же делать? Вводить бесконечную последовательную цепочку наблюдателей, расширяя нашу систему до бесконечности? Приписать нашему сознанию реальную материальную силу и способность к редукции системы или даже, как иногда предлагается, создать некую новую физику сознания [7]? Все гораздо проще и красивее.

В случае, когда кот находится под внешним наблюдением и наблюдатель или среда не включены в систему, как мы видели, парадокса не возникает. Проблема тут состоит в том, чтобы четко определить конкретный возможный вид внешнего воздействия, что точно невозможно, поскольку воздействующие объекты не включены в рассмотрение. Что делать когда все внешние факторы входят в систему, в том числе, когда наблюдателем будет сам кот? Эволюция и динамика системы будет четко различна для случая обратимой динамики без редукции и необратимой с редукцией. Наиболее ярко это иллюстрируется отсутствием возврата системы в исходное состояние при наличие редукции из-за возрастания энтропии в этом процессе и возвратами, причем почти периодическими, предсказываемыми обратимыми уравнениями квантовой механики. Разрешение этого парадокса заключается в том, что эта разница между двумя динамиками хоть реально и существует,но не может быть проверена на практике, т.е. экспериментально.Действительно, поскольку замкнутая изолированная система не может ни точно и однозначно померить свое полное состояние, ни решить систему уравнений описывающих их динамику, она не способна также ни полностью экспериментально проверить собственную динамику, ни сделать однозначный выбор между двумя возможными динамиками, как в данном конкретном случае, поскольку в проверяемой области они дают идентичный результат. Так система не может проверить относительно себя самой теорему Пуанкаре о возвратах в исходное состояне. Действительно, чтобы зафиксировать этот возврат, система должна "помнить" свое исходное состояние, чтобы сравнить его с текущим.Но полный возврат системы в исходное состояние исключает наличие такой "памяти".На самом деле, поскольку "память" является также частью системы она тоже должна вернутья в исходне состояние, т.е. просто будет стерта.Т.е., вернувшись в прежнее состояние, система должна неизбежно потерять память о всей своей истории и не сможет зарегистрировать возврат. Только внешний независимый наблюдатель с независимой внешней памятью способен на подобную проверку. Он же (или внешняя среда вместо него) и внесет неизбежные в квантовой механике погрешности в измеряемую систему, которые и объяснят редукцию. Таким образом, наличие двух различных динамик для одной и той же системы объясняется невозможность чисто экспериментально обнаружить разницу между ними, а не загадочными созидательными силами сознания.

Здесь важно отметить, что основой этого парадокса и причиной, что он столь явно проявился именно в квантовой механике, является то, что как раз в квантовой механике измерение неизбежно приводит, пускай, как правило, к малому , но конечному возмущению измеряемой системы. Именно наличие подобного возмущения, а не ее вероятностный характер, и отличает принципиально квантовую механику от классической, приводя к очень трудноразрешимым парадоксам.

Здесь стоит еще раз отметить, что, тем не менее, при строго определенных условиях, которые будут описаны ниже, точная проверка обратимых квантовых законов возможна, несмотря на это ограничение.

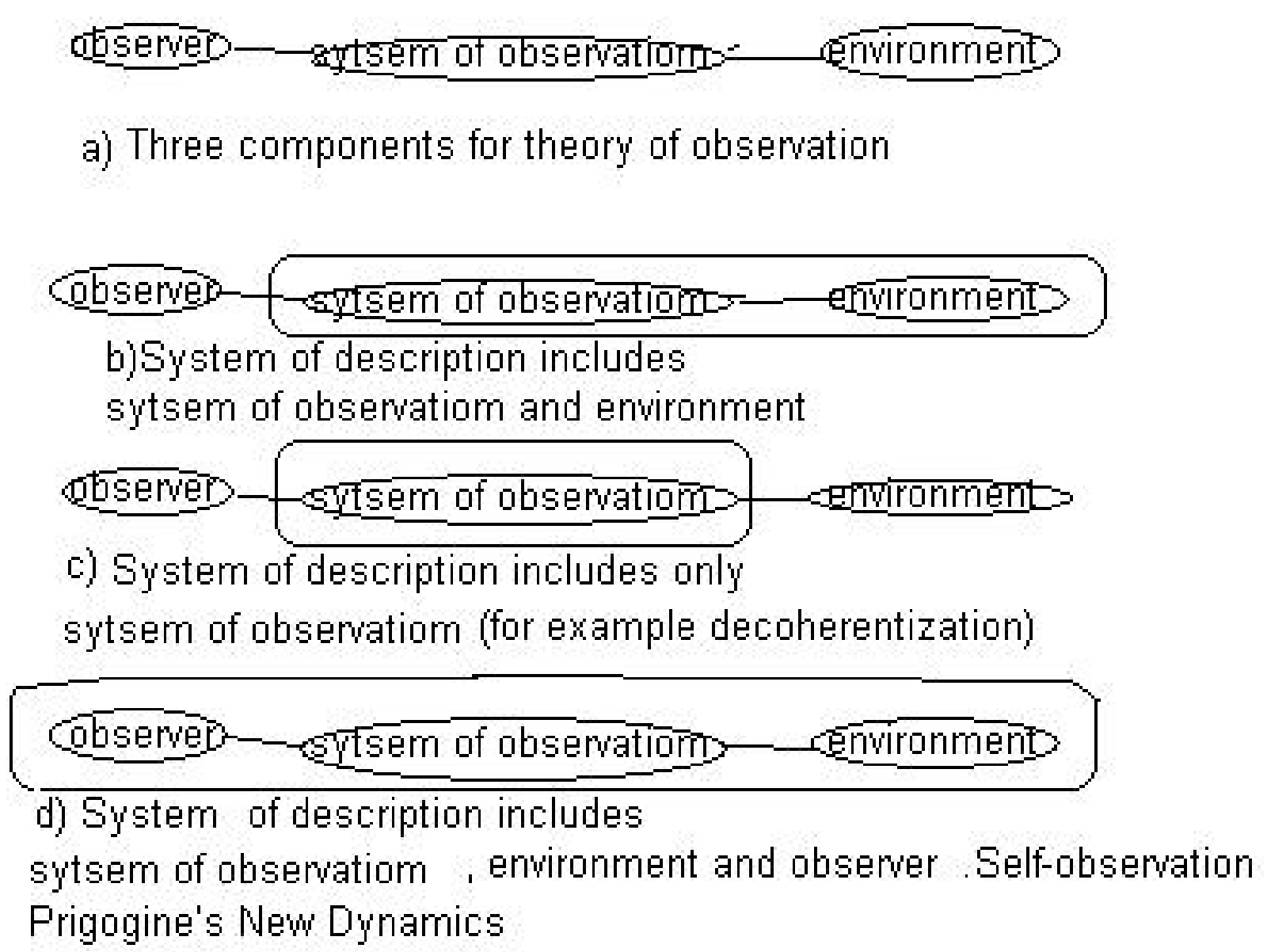

Figure 5 Interactions of three components for theory of observation.

**Часть 2. Классический аналог квантового парадокса «Шредингеровского кота».**

Парадокс «Шредингеровского кота», как правило, рассматривается, как чисто квантовый, не имеющий аналога в классической механике. Это, тем не менее, заблуждение. Аналог редукция волнового пакета имеет место и в классической механике! Действительно, закон возрастание энтропии ( в квантовой механике происходящий в момент редукция волнового пакета) происходит и в классической статистической механике, и по аналогии с квантовой механикой, входит в противоречие с обратимым характером законов движения и теоремой Пуанкаре о возвратах.Эти классические законы движения оставляют энтропию неизменной (если она определена через функцию плотности в фазовом пространстве), или она может как возрастать, так и убывать (если она определена через «огрубленную» функцию плотности в фазовом пространстве, т.е. усредненную в некоторой окрестности каждой точки функции плотности в фазовом пространстве). Чем же объясняется только возрастание энтропии в реальности? Введением Больцмановской «гипотезы молекулярного хаоса» [1], [2], которая является одним из видов «огрубления» функции плотности в фазовом пространстве Это эквивалентно пренебрежению корреляциями между скоростями и положением разных молекул, т.е. потере обратимости движения при обращении скоростей и нарушению теоремы Пуанкаре о возвратах в малую окрестность начального состояния. Эволюция системы становится необратимой.

Но мы здесь видим полную аналогию с редукцией волнового пакета! Она также приводит к потере корреляций (недиагональных элементов матрицы плотности) и необратимости движения! Т.е. «гипотеза молекулярного хаоса» или иной тип «огрубления» функции плотности в фазовом пространстве и является полным аналогом редукции в квантовой механике. Потеря корреляций при этих операциях эквивалентна обнулению недиагональных элементов матрицы плотности при редукции волнового пакета. Чем же можно объяснить введение этих дополнительных к классической механике и противоречащих им огрублений»? Да теми же причинами, что и редукция в квантовой механике: неразличимостью двух видов динамик(с и без огрублением) при реальных экспериментах (для замкнутых систем из-за невозможности «запоминания» начальных условий при самонаблюдении из-за возвратов, для внешнего же наблюдения взаимодействием наблюдателя или окружения с наблюдаемой системой) . Но тут мы и сталкиваемся с главным и принципиальным различием классической и квантовой механик.

Если в квантовой механике взаимодействие наблюдателя с наблюдаемой системой всегда присутствует и конечно, то в классической механике оно может быть теоретически сведено к нулю! На деле же оно всегда имеет место и конечно. Этим и объясняется противоречие между теоретической наблюдаемостью убывания энтропии и ее отсутствием в реальности в больших системах: реальное, конечное и малое взаимодействие с наблюдателем или просто «окружением» приводит к разрушению процессов с убыванием энтропии. Действительно, процессы с убыванием энтропии, в отличие от процессов с возрастанием энтропии,сильно неустойчивы по отношению к хаотическому внешнему воздействию, что приводит к их разрушению и синхронизации стрел времени между наблюдателем, наблюдаемой системой и их окружением, даже для малого взаимодействия. Стрела времени определяется в направлении возрастания энтропии. Введением этих стрел времени и пытались объяснить ранее закон возрастания энтропии. При этом появлялся законный вопрос: поскольку оба направления стрелы времени равновероятны (максимально вероятны лишь равновесные состояния без явно выраженного направления) почему в реальности все эти стрелы сонаправлены? Это считалось неразрешимой загадкой, решение которой необходимо искать чуть ли не глубинах происхождения Вселенной [28], в то время как решение очень просто и лежит лишь в реальном, пусть и малом взаимодействии всех подсистем, приводящем к всеобщей синхронизации всех стрел времени**.**

Надо отметить, что теоретическая возможность нулевого взаимодействия между системами в классической механике и привела к тому, что парадокс «Шредингеровского кота» проявился в явной и четкой форме лишь в рамках квантовой механики, в то время как он характерен и для классической механики при постулировании малого, но конечного взаимодействия между системами, которое всегда имеет место в реальном мире, за исключением некоторых очень тонких и искусственных ситуаций, создаваемых самими людьми в экспериментах. К этим ситуациям мы сейчас и перейдем. Предварительно лишь условимся, что в дальнейшем тексте называя системы «реальными», мы будем подразумевать наличие такого взаимодействия, хоть и малого, с окружением или наблюдателем, хотя в классической механике оно может быть и нулевым.

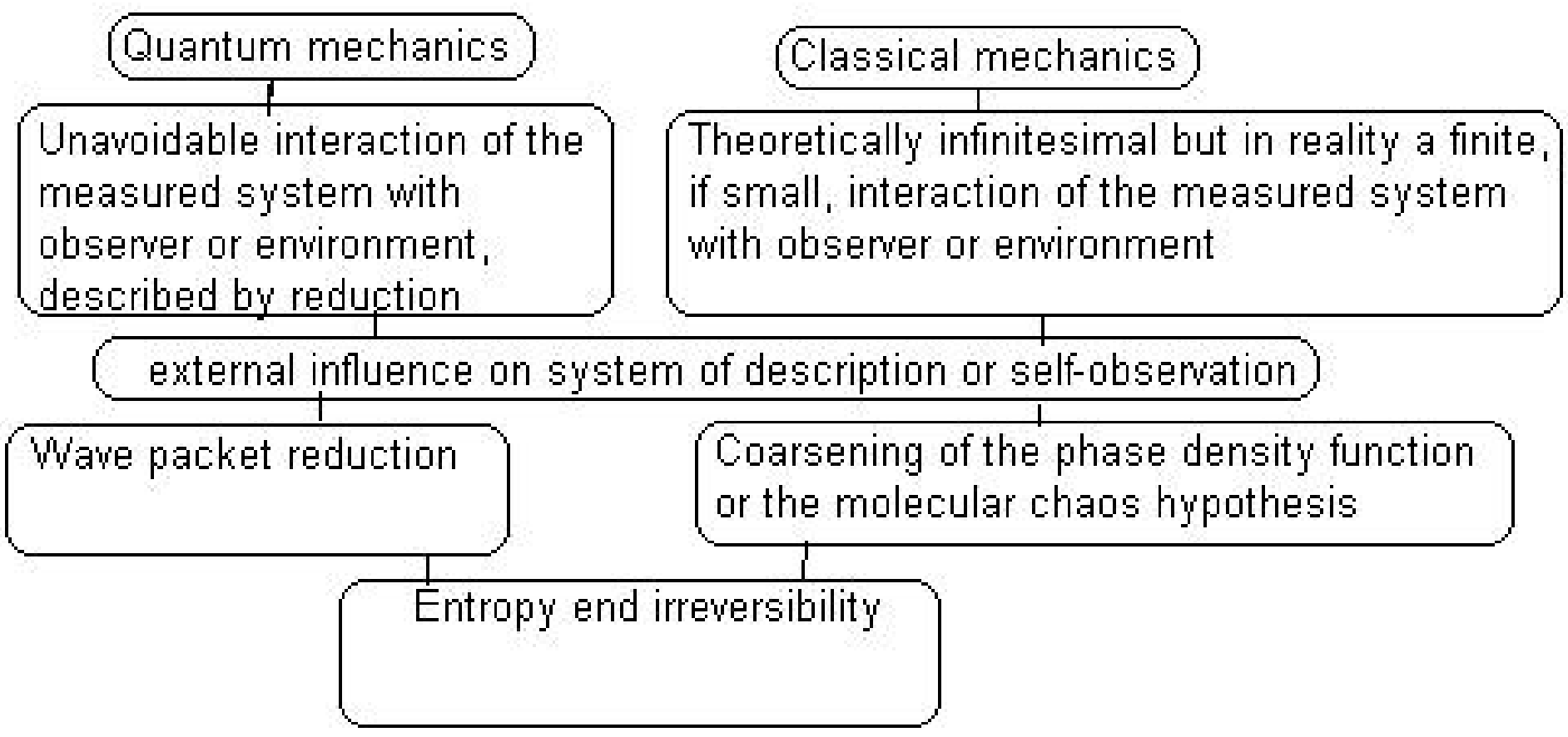

Figure 6 Sources of irreversibility and entropy in physics.

**Часть 3. Экспериментальное наблюдение квантового парадокса «Шредингеровского кота» и экспериментальная проверка Идеальной динамики.**

В этой Части мы обсуждаем, главным образом, системы конечного размера. Для бесконечной системы, например для звезд в бесконечной Вселенной, мы можем получить информацию о звездах, наблюдая свет, идущий от этих звезд. Такое измерение не влияет на сами звезды, поскольку этот свет не возвращается обратно к звездам после наблюдения в случае, если Вселенная бесконечна.

В классической механике такая ситуация создается очень просто: почти идеальной изоляцией системы от внешнего окружения и введением почти нулевого взаимодействия системы с наблюдателем. В квантовой механике это невозможно: измерение системы всегда связанно с неустранимым взаимодействием. Может показаться, что разница между редуцированной или не редуцированной динамикой ( будем не редуцированную динамику (в классической механике динамика до введения «огрубления») в дальнейшем называть Идеальной динамикой) становится в любом случае реально не проверяемой. Этот вывод, тем не менее, ошибочен. Измерение в квантовой механике возможно двух видов: когда результат измерения соответствует состоянию системы до или после измерения с теоретически 100% точностью. Оба состояния одновременно измерены быть не могут из-за взаимодействия с наблюдателем. Причем измерение, соответствующее состоянию системы после измерения является скорее не измерением, а «подготовкой» системы к измерению, поскольку исходное состояние меняется и остается неизвестным. Измерение, соответствующее состоянию системы после измерения будем именовать «наблюдением». Таким образом, квантовая система с «подготовкой» в начальный момент, почти полной изоляцией в промежуточном состоянии от наблюдения и внешней среды и «наблюдением» в конечном состоянии и является по сути полигоном для проверки Идеальной динамики, квантовым эквивалентом классической изолированой системы.

Подготовкой мы называем случай наблюдения ,когда резултаты измерения говорят о состоянии системы ПОСЛЕ измерения, а не до него.Наблюдение отвечает случаю , когда измрение определяет состояние системы ДО измерения. Само измерение – это взаимодействие , которое согласно законам квантовой механики неизбежно меняет состояние измеряемой системы, т.е. состояния ДО и ПОСЛЕ различны.
Квантовая система с «подготовкой» имеет следующие существенные недостатки:

1) исходное состояние меняется и остается неизвестным.
2) промежуточные состояния не измеряются и остаются неизвестны.По сути можно сравнить на соответствие Идеальной динамике лишь начальное и конечное состояния.
3) случаи, когда можно добиться столь полной изоляции редки и требуют огромных усилий.

Примеры таких систем:

1) Мезоскопические системы при низких температурах [10].
Эти системы за счет больших размеров близки к границе применимости закона больших чисел и почти макроскопичны. За счет низких температур(и, соответственно, импульсов и скоростей молекул) размеры квантового волнового пакета велики (в силу соотношения неопределенности) , близки к размеру системы и поддерживают, соответственно, квантовые корреляции. Взаимодействие с окружением слабо и его величина легко контролируется.Таким образом, возможна провека квантовых когерентных оссциляций или туннельного эффекта на относительно больших, почти макроскпических системах. Все эксперименты проведенные до сих пор подтверждают выполнение Идеальной динамики, а не редукции в промежуточных состояниях.

2) Системы вблизи фазовых переходов II рода. Для таких систем велика длинна корреляции, сравнимая с размерами системы.

3) Может быть, какие-то типы живых процессов или их примитивные возможные прототипы. Этому посвящена последняя часть статьи.

**Часть 4. Определение Реальной Динамики,Идеальной Динамики, Непредсказуемой Динамики и Макроскопического Состояния. «Новая Динамика» Пригожина** (Приложение F) **.**

Идеальной Динамикой мы условились называть эволюцию системы, описываемую исходными уравнениями квантовой и классической механик. В реальности, кроме небольшого количества случаев, описанных выше, из-за невозможности полного самоописания системы или неустранимого
взаимодействия с внешним наблюдателем (средой) Идеальная Динамика экспериментально непроверяема и поведение системы становится, строго говоря, становится просто не предсказуемо.Появляется Непредсказуемая Динамика вместо Идеальной Динамики.

На практике, однако, большинство систем хорошо описывается и предсказывается законами физики. Как же это удается?!

Существует два основных фактора, приводящих к непредсказуемости:

1) Невозможность полного самоописания, если наблюдатель и среда включены в систему описания.Это накладывает ограничение на точное знание начальных условий движения.

2) Неконтролируемое взаимодействие внешнего наблюдателя или внешней среды с описываемой системой. Это накладывает ограничение на точное знание уравнений движения из-за неконтролируемого внешнего шума.

Есть, однако, решение этих трудностей. Оно состоит в замене полного описания системы на сокращенное через введение макроскопических переменных, являющимися некими функциями микропеременных.При этом здесь это понятие толкуется очень широко, например, знание скоростей и положений всех молекул с любой, но конечной точностью является также макроописанием системы.

**Потрясающий факт**, но для подавляющего числа реальных систем почти всегда существует (и не один!) набор макропеременных, при котором уравнения их движения становятся в очень широком диапазоне внешних шумов или погрешностей начальных условий НЕЗАВИСИМЫМИ (или почти НЕЗАВИСИМЫМИ) от величины и конкретного вида этих шумов или погрешностей начальных условий в течение промежутка времени меньшего половины времени возврата для периодических или почти периодических систем и даже бесконечному для хаотических систем или систем с бесконечным числом частиц или бесконечным размером.

Т.е. эта Реальная Динамика не зависит от погрешностей или внешних шумов, а зависит лишь от свойств самой системы, как и исходная Идеальная Динамика. Существует как минимум две причины делающие Реальную Динамику устойчивой к шуму: статистический закон больших чисел и дискретность квантовых переходов, обеспечивающую устойчивость химических связей [11].

Тут возникает очень важный вопрос, как выбрать макропеременные. Именно требование этой независимости от шума и накладывает ограничение на возможный выбор макропеременных. Например, кроме пары состояний «Шредингеровского кота» живой и мертвый мы можем выбрать в квантовой механике их полусумму и разность. Почему в реальности выбор остается лишь за парой кот живой или мертвый? Именно потому, что эта пара устойчива к малому шуму со стороны макроскопического окружения, в то время как их полусумма или разность распадаются на них даже при очень малом внешнем шуме (теорема Данери-Лойнжера-Проспери[30], [31] Daneri A., Loinger A., Prosperi G. M.)[14].

Возможны и другие ограничения на макропеременные, связанные, например, с желанием уменьшить их число или сделать их поведение более детерминированным.

Другим важным свойством Реальной Динамики является неоднозначность выбора самой этой динамики и набора макропеременных, который она описывает. По сути, мы разрабатываем некие новые фундаментальные законы для данного нового уровня описания, лишь опираясь на точную Идеальную Динамику, которая становится уже точно экспериментально не проверяемой, но конечный и точный выбор этих законов во многом определяется лишь их удобством для нас.

Все это позволяет заменить Непредсказуемую Динамику на предсказуемую Реальную Динамику макропеременных, которая получается через введение «огрубления» или «редукции» или массы аналогичных методов. Следует особо подчеркнуть, что в большинстве случаев Реальная Динамика не просто некое приближение Идеальной Динамики (как часто толкуется), разница между ними хоть реально и существует, но она просто НЕ наблюдаема экспериментально! Т.е. вопрос какая из этих динамик более верна становится просто бессмысленным.

Для распада частицы или перехода с одного энергетического уровня на другой существует "парадокс котелка, который никогда не закипит". Он заключается в том, что если сделать интервалы между актами регистрации этого перехода достаточно малыми распад или переход вообще не происходит. Редукция волного пакета, происходящая в момент регистрации, меняет нормальный ход процесса при слишком частых регистрационных измерениях. Если же этот интервал времени выбрать достаточно большим (много больше t= E/h, E-разнитса енергии уровней , h –постоянная Планка) процесс распада почти не нарушается и почти не зависит от временного интервала между актами регистрации. Это и есть область "Реальной динамики".

В соответствии с двумя факторами непредсказуемости существует два типа Реальных Динамик:

1) Если наблюдатель и среда включены в систему описания.Это накладывает ограничение на точное знание начальных условий движения и приводит к Реальной Динамике не зависящей от них в широком интервале их значений. Это наиболее популярный тип Реальной  Динамики, поскольку такая Динамика носит  «объективный» характер и не  зависит от внешних  факторов,  хотя, на самом деле, оба типа Реальных Динамик определяются лишь параметрами самой системы. К этому типу и относится и широко известная «Новая Динамика»,  разработанная Пригожиным [9], [12].
2) Если наблюдатель или среда НЕ включены в систему описания. Неконтролируемое взаимодействие внешнего наблюдателя или внешней среды с описываемой системой накладывает ограничение на точное знание уравнений движения из-за неконтролируемого внешнего шума и приводит к Реальной Динамике не зависящей от этого шума в широком интервале его значения и вида. Соответствует широко используемой и применяемой в квантовой механике «декогеренизации» квантовых систем, взаимодействующих с внешними «большими» макросистемами.

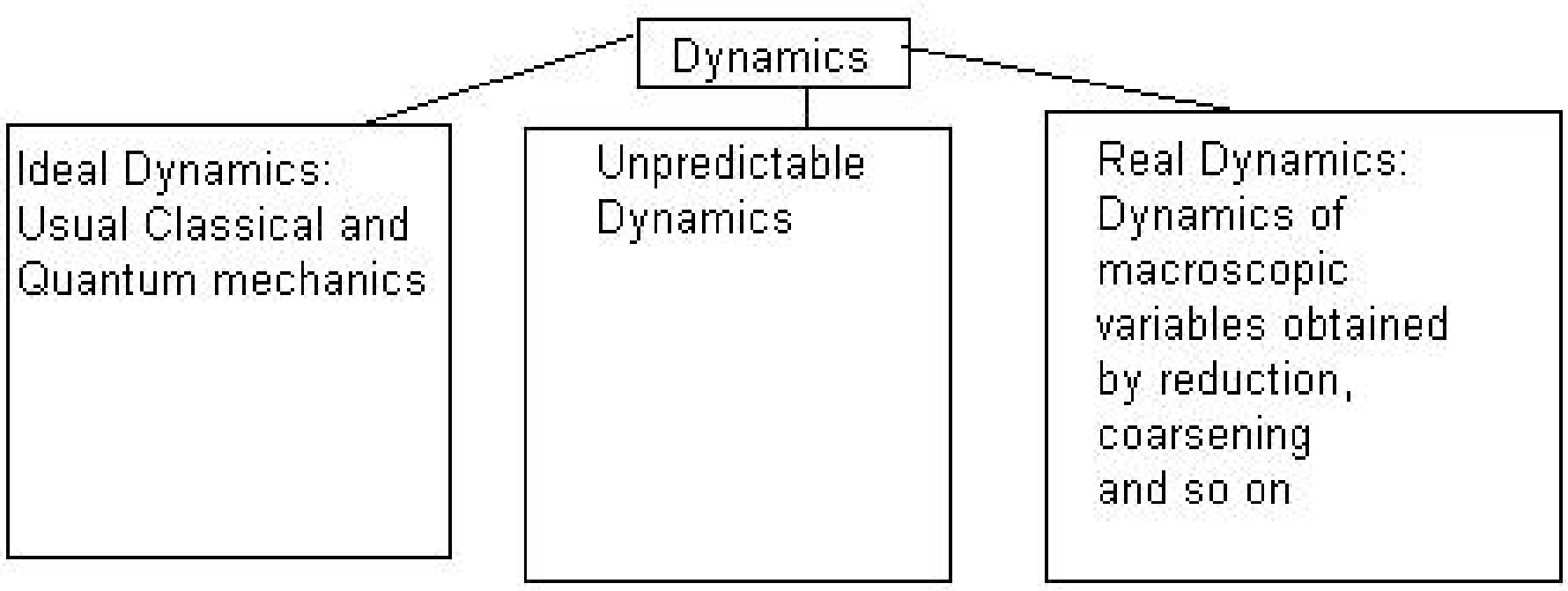

Figure 7 Three types of Dynamics.

Остановимся подробнее на «Новой Динамике», разработанной Пригожиным [9], [12] (Приложение F). Она отличается от других аналогичных методов удачным выбором процедуры «огрубления». Дело в том, что большинство реальных замкнутых систем в классической механике - это системы с перемешиванием, где подавляющая часть траекторий экспоненциально неустойчивы. Их аналогом в квантовом случае являются системы бесконечного размера или системы с бесконечным числом частиц.Эти системы и рассматривает теория Пригожина. За бортом остается хоть и ограниченный, но важный класс классических периодических и почти периодических систем и почти все квантовые замкнутые системы, которые почти всегда обладают тем же свойством. Казалось бы, приведенные выше квантовые и классические системы рассматриваемые  Пригожиным принципиально различны в этом плане.В квантовых бесконечных системах время возврата бесконечно, а в классических хаотических системах имеет хоть и случайную, но конечную величину. Однако из-за неизбежных погрешностей самоизмерения в классических системах начальные условия размазаны в малой окрестности и благодаря случайным величинам времен возврата полный возврат системы, рассматриваемый как не одна точка в фазовом пространстве, а вся ее малая окрестность возможен тоже лишь за бесконечное время. Для индивидуальных же (одна точка) систем эти случайные времена возврата не могут быть самонаблюдаемы, внешний же реальный наблюдатель всегда внесет свою корректирующую погрешность.

Функция фазовой плотности обладает свойством сохранения фазового объема первоначальной малой окрестности. Поскольку для систем с перемешиванием близкие траектории в одном направлении экспоненциально расходятся из сохранения фазового объема следует, что в другом направлении они должны столь же быстро сходиться. В этом направлении и предлагается делать огрубление. Его максимальная величина определяется именно условием независимости или очень слабой зависимости макропеременных от величины огрубления, как и следует делать в реальной динамике. Эта процедура огрубления обладает замечательным свойством, выделяющим ее среди других: уравнения движения для огрубленной или не огрубленной функции фазовой плотности остаются эквивалентными при этой процедуре, в том смысле, что она перестановочна с процедурой огрубления т.е. безразлично, что сделать вначале: провести процедуру огрубления и использовать Пригожинские уравнения для получения конечной огрубленной функции фазовой плотности или использовать Идеальную Динамику, а в конце сделать это же огрубление.

При обращении скоростей, площадь охватываемая неогрубленной функцией фазовой плотности не меняется. Для обычных методов огрубления это свойство сохраняется, нарушается лишь обратимость уравнений движения, правда, как уже много раз говорилось, не регистрируемая экспериментально в реальных ситуациях. В уравнениях же Пригожина необратимость появляется из-за несимметричности процедуры огрубления при обращении скоростей.

В фазовом пространстве скоростей и координат всех молекул, начальное положение системы соответствует компактной "капле". За счет экспоненциальной неустойчивости (условие "перемешивания") в одном направлении и сжатия в ортогональном направлении (что вытекает из закона сохранения фазового объема) эта "капля" расплывается, давая длинные тонкие "ветви".Направление этих "ветвей" соответствует направлению экспоненциального расширения. Идея "огрубления", высказанная Пригожиным, состоит в том, чтобы усреднять ("огрублять") функцию фазовой плотности не во всех направлениях, а лишь в направлении "сжатия". "Уширение" функции фазовой плотности в направлении сжатия практически не повлияет на динамику изменения этой функции во времени. При обращении скоростей (т.е. обращении времени) направление сжатия становится направлением расширения и динамика функции, полученной из огрубленной функции обращением скоростей, будет сильно отличаться от динамики функции, полученной из НЕогрубленной функции обращением скоростей, т.е. появляется желаемая "ассиметрия" времени.

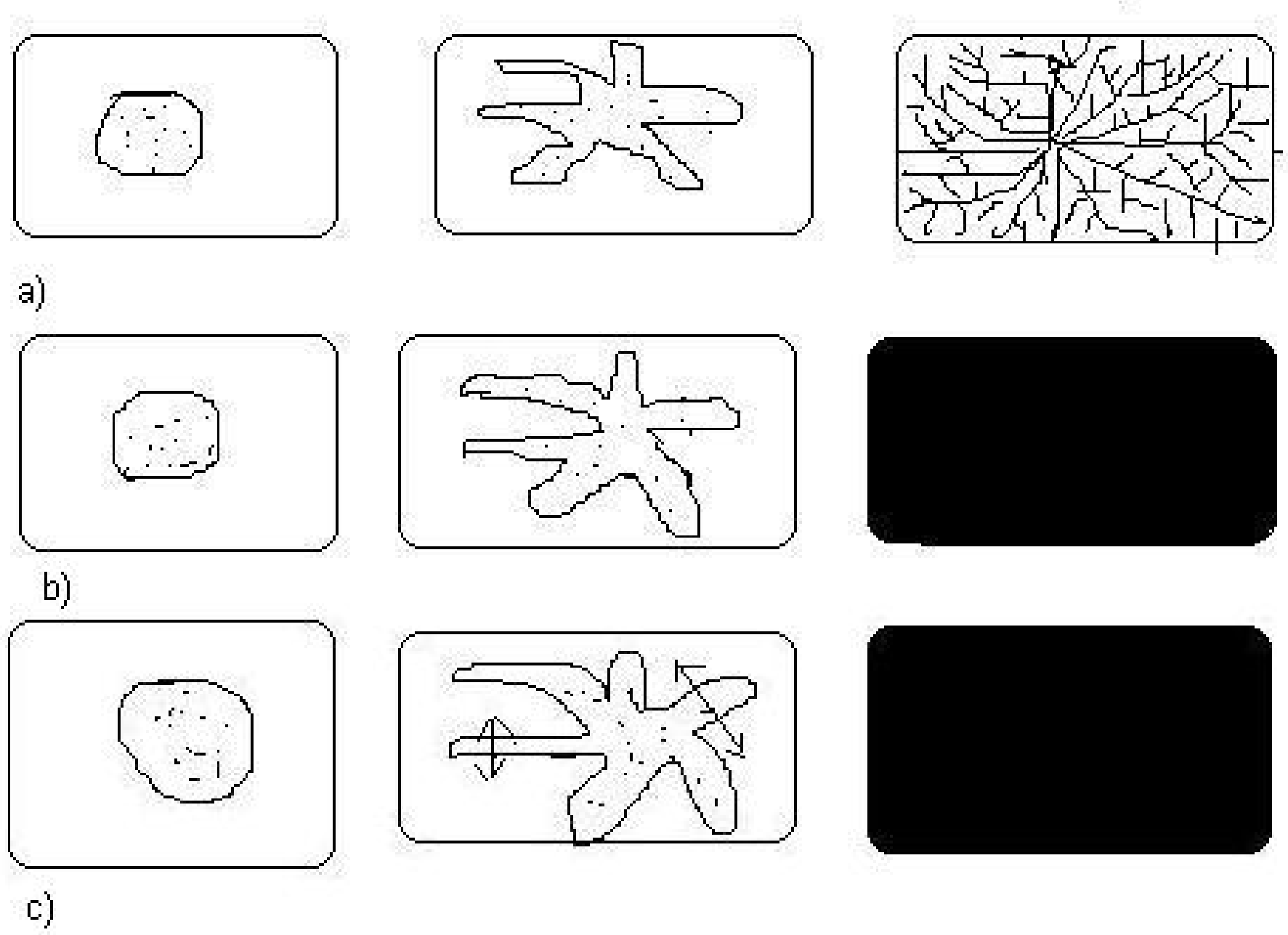

Figure 8a a)phase space "droplet" spreading. b)with isotropic "coarsening"
c)with anisotropic Prigogine "coarsening" . <-> is direction of "coarsening"

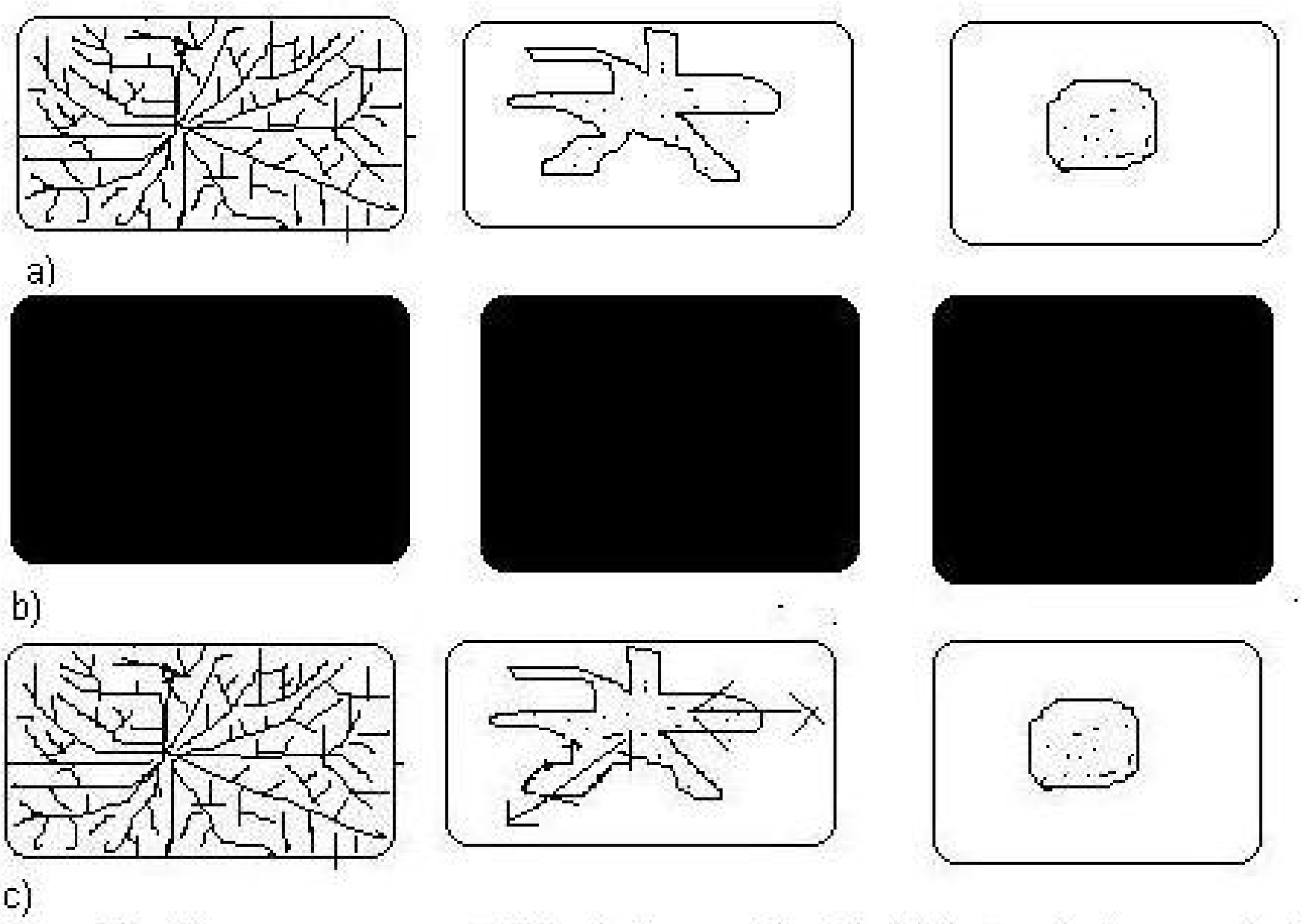

Figure 8b a)inverse process to "droplet" spreading b)with isotropic "coarsening"
c)with anisotropic Prigogine "coarsening" <-> is direction of "coarsening"

anisotropic Prigogine "coarsening" almost doesn't change dynamics.

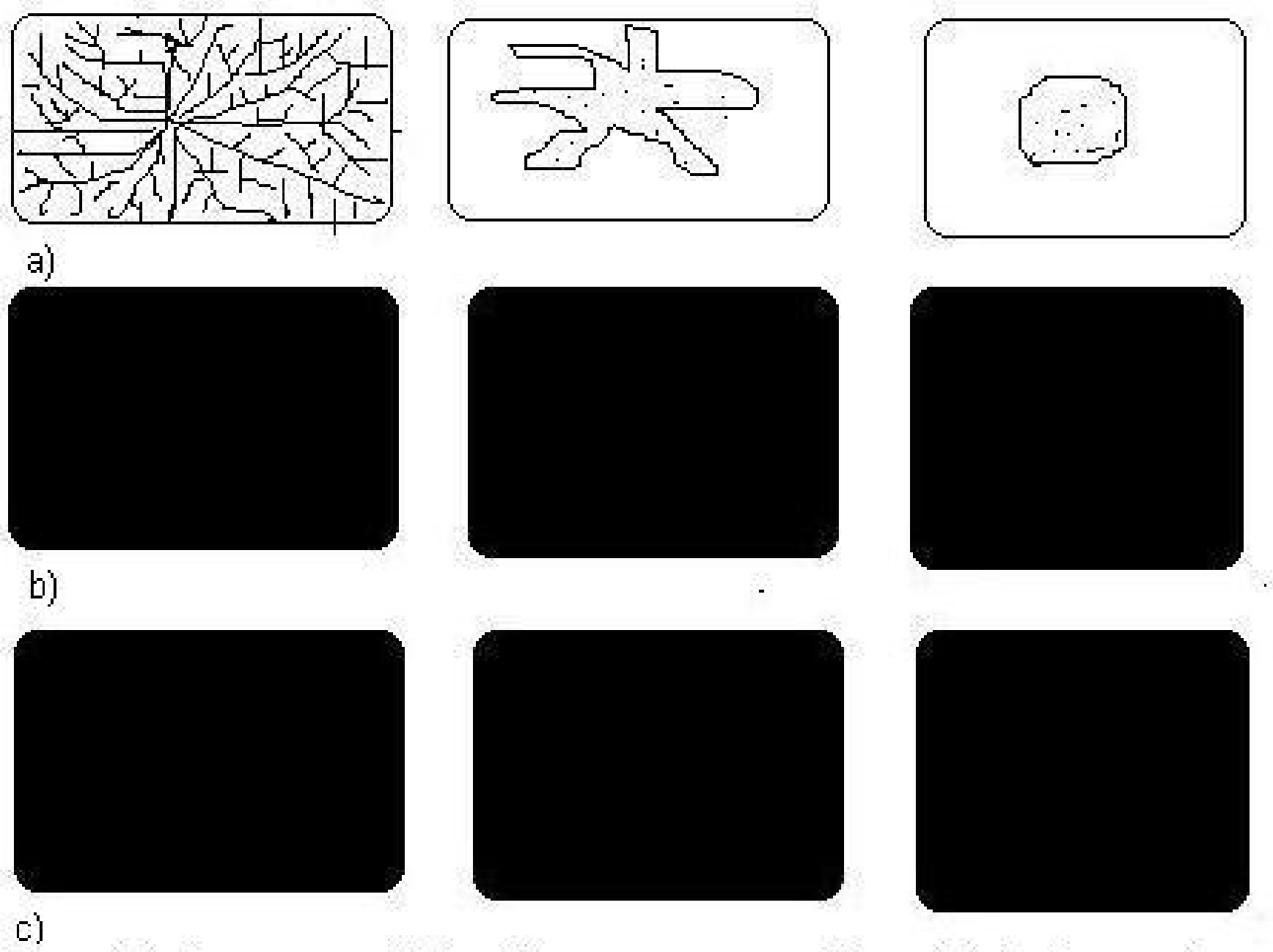

Figure 8c Processes obtained from processes on Figure 8a by inverse of molecules velocities in final state.Processes with "coarsening" are irreversible.

Энтропия функции фазовой плотности возрастает с увеличением фазового объема, охватываемого фазовой “каплей”. Следовательно, как можно видеть из рисунков, для неогрубленной функции энтропия не меняется, а для огрубленной возрастает. При этом для анизотропного огубления Пригожина динамика изменяется гораздо меньше, чем для изотропного, обычного огрубления.

Какова величина огрубления ? Такова, чтобы обеспечить реальную экспериментальную неразличимость «Новой» Реальной Динамики и Идеальной Динамики на протяжении бесконечного времени для описанных систем с перемешиванием. Это обстоятельство не было отмечено школой Пригожина.

Кстати, аналогичное огрубление может быть сделано и для периодических или почти периодических систем, правда, неразличимость Идеальная и «Новой» Динамики будет иметь место в этом случае лишь в течении времени равном половине периода возврата, что на самом деле и достаточно, как было показано выше, поскольку даже уже эти возвраты в реальных системах не наблюдаемы экспериментально. Также и это обстоятельство и не было замечено школой Пригожина. Какова величина огрубления? Такова чтобы обеспечить реальную экспериментальную неразличимость «Новой» Реальной Динамики и Идеальной Динамики на протяжении полупериода времени для описанных периодических или почти периодических систем с перемешиванием в течении только полупериода .

Много споров возникает в связи с вопросом, что истинно Идеальная Динамика или Реальная «Новая Динамика» Пригожина[28]? Этот спор очень похож на спор о том, что вокруг чего вращается Земля вокруг Солнца или наоборот? На самом деле, по самому определению движения выбор остается за нами и определяется лишь красотой и нашим удобством, аналогично тому, чем в математической науке определяется выбор определений и теорем, и теория этого выбора, кстати, является еще неоткрытым континентом ее основ, в отличие от теоремы Геделя! **(Примечание о т. Ферма: Кстати, результатом теоремы Геделя многие и объясняли труднодоказуемость т. Ферма, и итоговый результат оказался очень близок к этому утверждению. Т. Ферма получена как следствие гораздо более общей теоремы, чем она сама [13], и также включающей на порядок больше** **аксиом, чем теория натуральных чисел, на основе которых т. Ферма может быть сформулирована. Кстати сказать, было бы крайне любопытно получить от математиков не только 150 страничное доказательство, но и полный и вполне обозримый список аксиом математики, с указанием, какие из них использовались, а какие нет для этого доказательства. Аналогично этому, доказательство непротиворичивости арифметики получается введением трансфинитной индукции и соответственным расширением списка аксиом. Правда, тут отсутствует обобщенная теорема, новая аксиома трансфинитной индукции вводится «руками». Что считать аксиомой, а что теоремой, опять таки, дело красоты и удобства. А красота спасет мир!)**

Аналогично этому, разница между «Новой» и Идеальной Динамикой в большинстве реальных ситуаций экспериментально ненаблюдаема и выбор по большому счету произволен. В тех редких случаях «больших» систем, когда Идеальная Динамика точно проверяема она всегда пока что побеждает. Может быть исследуемые системы пока не достаточно большие? Ответ за дальнейшими экспериментами. Что окажется верным и какая Динамика из очень многих возможных одержит победу можно лишь гадать. Ситуация тут похожа на Великие Струнные Теории и Великие Объединения: покачто лишь догадки, а до эксперимента, который должен дать ответ может быть сотни лет, если не поможет Космос или Пришельцы, впрочем и в теории гравитации Энштейна, которая почему-то считается несомненной истиной (вот она,сила авторитета!), мы имеем пока в эксперименте лишь намеки на истину **(вспомним теории гравитации Мильгрома и Логунова и загадочное темное вещество)** . Сколько же ждать тут? Только Бог знает.

Закончим эту часть важным замечанием. Наряду с системами, описываемыми Реальной Динамикой или Идеальной Динамикой могут существовать системы, которые при попытке описать их подробно и детально отвечают Непредсказуемой Динамике, т.е. их детальное описание хоть и теоретически возможно, но не проверяемо экспериментально и не реализуемо на практике. Возможно к такому типу систем и относятся живые системы.

**Часть 5. Жизнь и смерть.**

Отметим с самого начала, что если предыдущие части носили более или менее строгий характер, данная в силу очевидных причин носит более гипотетический характер и является скорее набором гипотез.

Будем исходить в данной статье из положения, что жизнь полностью соответствует законам физики.

Следующие вопросы должны быть обсуждены ниже:

Что такое жизнь и смерть с точки зрения физики?

Есть ли у живой материи некие свойства не совместимые с физикой?

Чем живые системы отличаются от неживых с точки зрения физики?

Когда у живых систем появляется сознание и свобода воли
с точки зрения физики?

Жизнь определяется, обычно, как особая высокоорганизованная форма существования органических молекул, обладающая способностью к обмену веществ, размножению, адаптации, движению, реакцией на внешние раздражители, способностью к самосохранению в течении долгого времени или даже повышению уровня самоорганизации. Это верное, но слишком узкое определение: многие из живых систем обладают лишь частью из этих свойств, некоторые из них присущи и неживой материи, вполне возможны и неорганические формы жизни.

Первую попытку описать жизнь с точки зрения физики дал Шредингер [11]. В своей работе он определил жизнь как апериодический кристалл, т.е. высокоупорядоченную (и, соответственно, обладающую низкой энтропией и «питающуюся» негоэнтропией из окружения, т.е. принципиально открытую систему), но не основанную на простом повторении, в отличие от кристалла, форму материи и привел две причины делающие Реальную Динамику живых систем устойчивой к их внутреннему и внешнему шуму: статистический закон больших чисел и дискретность квантовых переходов, обеспечивающую устойчивость химических связей. Сам принцип действия живых организмов он уподобляет часам: и там и там возникает «порядок из порядка» несмотря на высокую температуру.

В своей работе советский биофизик Бауэр [12] определил, что не только высокая упорядоченность (и, соответственно, низкая энтропия) проявляются не только в неравновесности распределения веществ в живой материи, но и сама структура живой материи является низкоэнтропийной и сильно неустойчивой. Эта неустойчивая структура не только поддерживается за счет процесса обмена веществ, но и является их катализатором. Это предположение верно лишь частично, например белки или вирусы сохраняют свою структуру и в кристаллической форме, но их низкоэнтропийные и сильно неустойчивые модификации и сочетания внутри живой материи обладают этим свойством. С течением времени, тем не менее, происходит постепенная деградация структуры, что и приводит к неизбежности смерти и необходимости размножения для сохранения жизни. Т.е. процесс обмена веществ лишь очень сильно замедляет распад сложной структуры живой материи, а не поддерживает ее все время неизменной. Экспериментальные результаты, приведенные Бауэром, подтверждают выделение энергии и соответственно увеличение энтропии в процессе автолиза, т.е. распада живой материи. Он видит его причину на первой стадии процесса в неустойчивости самой исходной структуры без поддерживающего ее обмена веществ и на второй стадией процесса в действии протолитических (разлагающих) ферментов, освобождающихся или появляющихся при автолизе. Наличие этой избыточной структурной энергии Бауэр и считал неотемлимой характеристикой жизни.

Во всех этих работах дано рассмотрение отдельного живого организма, в то время как жизнь, как совокупность всех организмов в целом (биосфера) может быть рассмотрено и определено. Сюда же относится и вопрос о происхождении и источнике жизни. Наиболее полный и современный ответ на эти вопросы с точки зрения физики был дан в работе Элицура [18]. В ней он рассматривает источник жизни как ансамбль саморазмножающихся молекул. Проходя через сито Дарвиновского естественного отбора, жизнь накапливает в своих генах информацию (или скорее знания в определении, данном Элицуром) об окружающей среде, повышая, тем самым, уровень своей организации (негоэнтропии) в соответствии со вторым началом термодинамики. Ламаркизм в его слишком прямолинейной формулировке приведен в противоречие с этим законом физики.Взгляд на широкий спектр работ в этой области отражен в статье. К недостаткам работы относятся:

1) Рассмотрение верно для жизни в целом, как явления, но не для отдельно взятого живого организма.
2) Предложенное доказательство отвергает лишь слишком грубую, прямолинейную модель Ламаркизма, в то время как есть много гипотез и опытов, иллюстрирующих возможность реализации его элементов даже в реальной жизни[32]. Идея Ламаркизма заключается в том, что полезные изменения, происходящие в организме в течении его жизни, запоминаются и передаются потомству. Это противоречит Дарвиновской концепции отбора, когда случайные изменения происходят в наследственном веществе и удачные из них закрепляются у потомков естественным отбором, а неудачные отбрасываются.
3) За самоорганизующимися диссипативными системами, предложенными Пригожиным, например ячейками Бернара, отрицается свойство адаптации, в

отличие от живых организмов. Естественно, их адаптивные способности не сравнимы с живыми системами, но в зачаточной форме, тем не менее, существуют. Так, например, ячейки Бернара,меняют свою геометрию или даже исчезают, как функция разницы температур между нижним и верхним слоем жидкости.Это и есть примитивная форма адаптации.

Равновесный ансамбль, находящийся в равновесии с термостатом, в квантовой механике в энергетическом представлении описывается диагональной матрицей плотности. Аналогично этому, в классической механике, в равновесии отсутствуют корреляции между молекулами, являющимися аналогами недиагональных элементов матрицы плотности.

Таким образом нарушение равновесия проявляется двояко: в неравновесном распределении диагональных элементов и в неравенстве нулю недиагональных элементов (что соответствует ненулевым корреляциям в классической механики), причем эти корреляции много более неустойчивы и гораздо быстрее затухают, чем отклонение диагональных элементов от равновесных величин.

Поскольку жизнь определена Бауэром как самоподдерживающаяся за счет движения и обмена веществ сильная неустойчивость, мы можем предположить, что большая часть этой неустойчивости проистекает из этих сильно неустойчивых корреляций (в квантовой механике недиагональных элементов), которые живые системы стремятся поддержать и сохранить в течении времени много большем их времени релаксации.

В неживых системах это достигается просто изоляцией системы, в живых же открытых системах, активно взаимодействующих с окружением это достигается их внешним и внутренним движением, метаболизмом и активным взаимодействием с окружающей средой.. Следует отметить, что живые системы поддерживают корреляции как между внутренними элементами, так и корреляции с окружением.

Введем понятие псевдоживых физических систем. Будем называть таковыми простые физические системы иллюстрирующие в зачаточной форме некие действительные или предполагаемые свойства живых систем. Так, рост кристаллов моделирует способность живых систем к размножению. Кстати, анализ этих систем позволяет найти слабое место в аргументах Вигнера [6] , [27], видящем противоречие между способностью к размножению и квантовой механикой.

Другой пример - это квантовые изолированные системы, демонстрирующие свойство сохранения корреляций, аналогичных поддержанию сильной неустойчивости в живых системах,связанной с сохранением корреляций или недиагональных элементов матрицы плотности.

Правда, это сохранение пассивно. Активное замедление времен релаксации недиагональных элементов матрицы плотности, более близкое к методам поддержания корреляций в живых системах, достигается в таких открытых системах, таких как микромазеры [25], которые являются еще одним примером псевдоживых систем. Диссипативные системы иллюстрируют свойства открытых живых систем к поддержанию низкой энтропии и примитивной адаптации к изменению условий окружающей среды.

Кстати, определение жизни, как систем, способствующих сохранению корреляций в противовес внешнему шуму, хорошо объясняет загадочное молчание КОСМОСА, т.е. отсутствие сигналов от других разумных миров. Вселенная произошла из единого центра (Большой Взрыв) и все ее части коррелированны, жизнь лишь поддерживает эти корреляции и существует на их основе. Поэтому процессы возникновения жизни в различных частях скоррелированны и находятся на одном уровне развития, т.е. сверхцивилизаций, способных достичь Земли пока просто нет. Эффектами дальних корреляций можно объяснить и часть поистине чудесных проявлений человеческой интуиции и парапсихологических эффектов. Причем здесь необязательна квантовая механика, подобные корреляции присущи и классической механике, имеющей аналоги недиагональных элементов матрицы плотности. Подобные корреляции часто ошибочно увязывают лишь с квантовой механикой.

Следующий вклад в понимание жизни сделал Бор [26]. Он обратил внимание, что полное измерение состояния системы вносит в квантовой механике неизбежные искажения в поведение системы, чем возможно и объясняется принципиальная непознаваемость жизни. Критика этих взглядов Бора Шредингером [19] не состоятельна. Она основана на том, что полное измерение состояния системы возможно и в квантовой механике, просто оно отлично от классического - оно вероятностно. Проблема не в том, что такое измерение невозможно. Истинная проблема заключается в том, что подобное измерение меняет дальнейшее поведение системы, в отсутствии измерения оно было бы иным [20]. Измерение нарушает тонкие корреляции между частями системы, меняя ее поведение. Это относится не только к квантовой механике, но и к классической механике, где между реальными системами существует конечное взаимодействие.

Псевдоживыми системами иллюстрирущими свойство измерения нарушать динамику систем являются оссцилирующие квантовые почти изолированные системы, изменяющиеся по схеме: A -> сумма A и B -> B -> разница A и B-> A, где A и B - состояния системы. Измерение в каком состоянии находится система A или B нарушает состояния их суммы или разницы, меняя реальную динамику системы и уничтожая корреляции ( не диагональные элементы матрицы плотности) между A и B в этих состояниях [10].

Успехи молекулярной генетики не опровергают этой точки зрения. Построение Реальной Динамики жизни в принципе возможно. Действительно, живые системы - это открытые системы, активно взаимодействующие со случайным окружением. Внешний наблюдатель взаимодействует с ними обычно много слабее и не может вызвать принципиальное изменение в их поведении. Однако попытка понять и предсказать жизнь слишком подробно и детально нарушит сложные и тонкие корреляции, сохраняемые жизнью, и приведет к Непредсказуемой Динамике живых систем и давая эффект, предсказанный Бором. Возможно, особо тонкая человеческая интуиция и некоторые парапсихологические эффекты и лежат в этой области Непредсказуемости. То что они могут лежать только в этой узкой области на грани постижимости точной наукой и не дает естественному отбору усилить эти свойства, так и не дает нам возможности постичь полностью эти явления средствами науки [21], [22].

**Часть 6. Заключение.**

Приведенная статья не является лишь философским абстрактным построением. Не понимание ее основ приводит к ошибкам. Подавляющее большинство реальных систем не описывается идеальными уравнениями квантовой или классической механики. Из-за всегда существующего ( и неизбежно существующего для измерений в квантовой механике) воздействия измерительной аппаратуры и окружения системы на эту систему эти уравнения нарушаются. Попытка включить измерительные приборы и окружение в описываемую систему приводит к самонаблюдаемой системе(Приложение G) . Такая система не может измерить и запомнить в полной мере собственные состояния и ее даже приближенное самоописание имеет смысл лишь для промежутков времени много меньших времени возврата, определяемого согласно теореме Пуанкаре, после которого, вся память о прошлом неизбежно стирается. Однако систему можно описать в этой ситуации не Идеальной, а Реальной динамикой, которая возможна, поскольку, для широкого интервала внешних шумов свойства этой динамики не зависят от величины и вида этих шумов, а определяются лишь свойствами самой системы. Реальная динамика может быть получена и на основе огрубления функции распределения или матрицы плотности, поскольку начальное состояние точно не определено. Разница между Реальной и Идеальной динамикой не может быть проверена даже в случае если наблюдатель и окружение включены в описываемую этой динамикой системы, поскольку самонаблюдение ограничено по точности и времени наблюдения. Так, например, возвраты системы, вытекающие из теоремы Пуанкаре для идеальной динамики, не могут быть наблюдаемы самой этой системой из-за эффекта стирания памяти о прошлых состояниях. Введение этой реальной динамки разрешает все известные парадоксы классической и квантовой механики.

Приведем несколько конкретных примеров ошибок, сделанных из-за непонимания этих основ сделанных, например, в теории полюсов для задачи движения фронта пламени и роста «пальца» на поверхности раздела жидкостей.

Севашинский[23] утверждал, что Идеальная Динамика полюсов приводит к ускорению фронта пламени, и это ускорение не вызвано шумом, поскольку оно не меняется при уменьшении шума и зависит лишь от свойств самой системы. Но ведь также Реальная Динамика, связанная с шумом, не зависит от него в широком интервале значений.

Танвир [24] нашел различие в росте «пальца» в теории и численных экспериментах, не поняв, что эта разница связана с численным шумом, приводящим к новой Реальной Динамике. Это лишь два рядовых примера, взятых из повседневной практики автора статьи, а встретить их можно много. Изложенные в этой статье результаты необходимы для понимания основ нелинейной динамики, термодинамики и квантовой механики.

**Литература.**

**Приложение А. Матрица плотности**

Рассмотрим пучок из $N_a$ частиц , приготовленных в состоянии $|\chi_a\rangle$ , и еще один независимый от первого пучок из $N_b$ частиц, приготовленных в состоянии $|\chi_b\rangle$ . Для описания объединенного пучка введем оператор ρ смешанного состояния, определяемый выражением

$$\rho=W_a|\chi_a\rangle\langle\chi_a| + W_b|\chi_b\rangle\langle\chi_b|$$

где $W_a=N_a/N$, $W_b=N_b/N$, $N=N_a+N_b$

Оператор ρ называют оператором плотности или статистическим оператором. Он описывает способ приготовления пучков и тем самым содержит всю информацию о полном пучке. В этом смысле смесь полностью определена своим оператором плотности. В частном случае чистого состояния $|\chi\rangle$ оператор плотности дается выражением

$$\rho=|\chi\rangle\langle\chi|.$$

Обычно более удобно записывать оператор ρ в матричной форме. Для этого выберем набор базисных состояний ( обычно $|+1/2\rangle$ и $|-1/2\rangle$) и разложим состояния $|\chi_a\rangle$ и $|\chi_b\rangle$ по этому набору согласно соотношению

$$|\chi_a\rangle=a_1^{(a)}|+1/2\rangle+a_2^{(a)}|-1/2\rangle,$$
$$|\chi_b\rangle=a_1^{(b)}|+1/2\rangle+a_2^{(b)}|-1/2\rangle.$$

В представлении состояний $|\pm1/2\rangle$ :

$$|\chi_a\rangle=\begin{pmatrix} a_1^{(a)} \\ a_2^{(a)} \end{pmatrix}$$

$$|\chi_b\rangle=\begin{pmatrix} a_1^{(b)} \\ a_2^{(b)} \end{pmatrix}$$

а для сопряженных состояний -

$$\langle\chi_a|=(a_1^{(a)*} , a_2^{(a)*} ) ,$$
$$\langle\chi_b|=(a_1^{(b)*} , a_2^{(b)*} ) .$$

Применяя правила умножения матриц, получим для «внешнего произведения»

$$|\chi_a\rangle\langle\chi_a|=\begin{pmatrix} a_1^{(a)} \\ a_2^{(a)} \end{pmatrix}(a_1^{(a)*} , a_2^{(a)*} )=\begin{pmatrix} |a_1^{(a)}|^2 & a_1^{(a)}a_2^{(a)*} \\ a_1^{(a)*}a_2^{(a)} & |a_2^{(a)}|^2 \end{pmatrix}$$

и аналогичное выражение для произведения $|\chi_b\rangle\langle\chi_b|$ . Подставляя эти выражения в оператор плотности, находим

матрицу плотности

$$\rho=\begin{pmatrix} W_a\left|a_1^{(a)}\right|^2+W_b\left|a_1^{(b)}\right|^2 & W_a a_1^{(a)}a_2^{(a)*}+W_b a_1^{(b)}a_2^{(b)*} \\ W_a a_1^{(a)*}a_2^{(a)}+W_b a_1^{(b)*}a_2^{(b)} & W_a\left|a_2^{(a)}\right|^2+W_b\left|a_2^{(b)}\right|^2 \end{pmatrix}$$

Поскольку при выводе этого выражения были использованны базисные состояния $|\pm1/2\rangle$, полученное выражение называют матрицей плотности в $\{|\pm1/2\rangle\}$-представлении.

В заключение несколько слов о статистической матрице $P_0$,
обладающей замечательными свойствами. Мы знаем, что в классической статистической термодинамке все возможные макроскопические состояния системы рассматриваются как априори равновероятными (другим словами, они считаются равновероятными, если нет каких-либо сведений о значении полной энергии или о контакте с термостатом, поддерживающим постоянную температуру системы, и т.д.). По аналогии с этим в волновой механике все состояния системы, определяемые различными функциями, образующие полную систему ортонормированных функций, можно априори предполагать равновероятными. Пусть $\varphi_1,\ldots,\varphi_k$,- такая система базисных функций $\varphi_k$; зная, что система характеризуется смесью состояний $\varphi_k$, в отсутствие какой-либо другой информации можно считать, что статистическая матрица системы имеет вид

$P_0=\sum_k pP_{\varphi_k}$ , где $\sum_k p = 1$ ,

т.е., что $P_0$ - статистическая матрица такого смешанного состояния, для которого все веса равны между собой. Принимаем $\varphi_k$ за базисные функции, матрицу $P_0$ можно представить в виде

$(P_0)_{kl}=p\delta_{kl}$

Если статистическое состояние ансамбля систем в начальный момент времени характеризуется матрицей $P_0$ и если во всех системах ансамбля провести измерение одной и той же величины А, то статистическое состояние ансамбля будет по прежнему характеризоваться матрицей $P_0$.

Уравнения движения для матрицы плотности ρ

$$i\frac{\partial\rho_N}{\partial t} = L\rho_N$$

где L - линейный оператор:

$L\rho=H\rho-\rho H=[H,\rho]$,

где H - оператор энергии системы.

Если А – оператор некой наблюдаемой величины,

Среднее значение этой величины может быть найдено следующим образом:

$\langle A\rangle=\mathrm{tr}A\rho$

**Приложение B** Редукция матрицы плотности и теория измерения.

Пусть при измерении некоторого объекта мы «четко различаем» состояния $\sigma^{(1)}$, $\sigma^{(2)}$, ... . Производя измерения над объектом, находящимся в этих состояниях, мы получаем числа $\lambda_1$, $\lambda_2$, ... . Начальное состояние измерительного прибора обозначим через a. Если измеряемая система сначала находилась в состоянии $\sigma^{(\nu)}$, то состояние полной системы «объект измерения плюс измерительный прибор» до того, как они вступили во взаимодействие будет определяться прямым произведением $a \times \sigma^{(\nu)}$ . После измерения:

$$a \times \sigma^{(\nu)} \rightarrow a^{(\nu)} \times \sigma^{(\nu)}$$

Пусть теперь начальное состояние будет не «четко различимым», а произвольной комбинацией $\alpha_1 \sigma^{(1)} + \alpha_2 \sigma^{(2)}+\ldots$

таких состояний. В этом случае из линейности квантовых уравнений следует

$$a \times [\Sigma\alpha_\nu\sigma^{(\nu)}] \rightarrow \Sigma\alpha_\nu[a^{(\nu)} \times \sigma^{(\nu)}]$$

В итоговом состоянии, возникающем в результате измерения,

существует статистическая корреляция между состоянием объекта и состоянием прибора: одновременное измерение у системы «объект измерения плюс измерительный прибор» двух величин ( первой- подлежащий измерению характеристики исследуемого объекта и второй-положение стрелки прибора) всегда приводит к согласующимся результатам. Вследствие этого одно из названных измерений становится излишним - к заключению о состоянии объекта измерения можно прийти на основании наблюдения за прибором.
Результатом измерения является не вектор состояния, представляемый в виде суммы, стоящей в правой части найденного соотношения, а так называемая смесь, т.е. один из векторов состояния вида

$$a^{(\nu)} \times \sigma^{(\nu)}$$

и что при взаимодействии между объектом измерения и измерительным прибором это состояние возникает с вероятностью $|\alpha_\nu|^2$. Данный переход называется редукцией волнового пакета и соответствует переходу матрицы плотности из недиагонального $\alpha_\nu\alpha_\mu^*$ в диагональный вид $|\alpha_\nu|^2\,\delta_{\nu\mu}$. Этот переход не описывается квантовыми уравнениями движения.

**Приложение C** Функция фазовой плотности

В данный момент времени t состояние системы N одинаковых частиц с точки зрения классической механики определяется заданием значений координат $\mathbf{r}_1, \ldots, \mathbf{r}_N$ и импульсов $\mathbf{p}_1, \ldots, \mathbf{p}_N$ всех N частиц системы. Для краткости будем использовать обозначение

$$x_i=(\mathbf{r}_i,\mathbf{p}_i) \quad (i=1, 2, \ldots, N)$$

для совокупности значений координат и компонент импульса отдельной частицы и обозначение

$$X= (x_1, \ldots, x_N)=(\mathbf{r}_1, \ldots, \mathbf{r}_N, \mathbf{p}_1, \ldots, \mathbf{p}_N)$$

для совокупности значений координат и значений импульсов всех частиц системы. Такое пространство 6N переменных называют 6N-мерным фазовым пространством.
Чтобы определить понятие функции распределения состояний системы, рассмотрим набор одинаковых макроскопических систем - ансамбль Гиббса. Для всех этих систем условия опыта одинаковы. Поскольку, однако, эти условия определяют состояние системы неоднозначно, то разным состояниям ансамбля в данный момент времени t будут соответствовать разные значения X.
Выделим в фазовом пространстве объем dX около точки X. Пусть в данный момент времени t в этом объеме заключены точки, характеризующие состояния dM систем ансамбля из их полного числа M. Тогда предел отношения этих величин

$$\lim_{m\to\infty} dM/M=f_N(X,t)dX$$

и определяет фазовую функцию плотности распределения
в момент времени t.

$$\int f_N(X,t)dX=1$$

Уравнение Лиувилля для функции фазовой плотности
Можно записать в виде

$$i\frac{\partial f_N}{\partial t} = Lf_N = \{H, f_N\}$$

где L - линейный оператор:

$$L = -i\frac{\partial H}{\partial p}\frac{\partial}{\partial x} + i\frac{\partial H}{\partial x}\frac{\partial}{\partial p},$$

где H - энергия системы.

**Приложение D Огрубление функция фазовой плотности и гипотеза молекулярного хаоса**.

Огрублением функции плотности будем называть ее замену на приближенную величину.

Например

$f_N^*(X,t)=\int_{(y)} g(X-Y) f_N(Y,t)$

где

$g(X)=1/\Delta\ D(X/\Delta)$

$D(x)= 1$ for $|X|<1$
$D(x)= 0$ for $|X|\geq 1$

Другой пример огрубления - «гипотеза молекулярного хаоса»

Замена двухчастичной функции распределения на произведение одночастичных функций
$f(x_1,x_2,t) \to f(x_1,t)f(x_2,t)$

**Приложение E Определения энтропии.**
В качестве определения энтропии укажем выражение

$S=-k \int_{(X)} f_N(X,t) \ln f_N(X,t)$

Для квантовой механики через матрицу плотности

$S=-k\ \mathrm{tr}\ \rho \ln \rho$ [29]

где tr - след матрицы.

Эти энтропии не меняются при обратимой эволюции. Чтобы получить меняющиеся энтропии в качестве $f_N$ или $\rho$ используются их огрубленные величины.

**Приложение F Новая динамика Пригожина.**
В статье неоднократно упоминается "Новая Динамика" Пригожина. Дадим здесь ее очень краткое изложение, опираясь на книги [9, 12]
Позвольте ввести линейный оператор $\Lambda$ на функции $\rho$ фазовой плотности или матрицы плотности

$\tilde{\rho} = \Lambda^{-1}\rho$
$\Lambda^{-1}\ 1=1$
$\int \tilde{\rho} = \int \rho$
$\Lambda^{-1}$ сохраняет положительность.
Такой, что функция $\tilde{\rho}$ обладает свойством:для функции $\Omega$ определенной как
$\Omega = tr\tilde{\rho}^+\tilde{\rho}$ или $\Omega = tr\tilde{\rho}\ \ln\tilde{\rho}$
имеем $d\Omega/dt \leq 0$
Уравнение движения для $\rho$

$$\frac{\partial\tilde{\rho}}{\partial t} = \Phi\tilde{\rho}$$

где $\Phi = \Lambda^{-1} L\Lambda$

Ф-необратимая марковская полугруппа

$\Lambda^{-1}(L)=\Lambda^{+}(-L)$

Для функции фазовой плотности оператор $\Lambda^{-1}$ соответствует
огрублению в направлении сжатия фазового объема. Для квантовой механики такой оператор может быть найден лишь для бесконечного объема или бесконечного числа частиц. Введем в квантовой механике проекционный оператор P, обнуляющий недиагональные элементы матрицы плотности. Оператор Ф и базисные вектора матрицы плотности выбираются таким образом, чтобы сделать оператор Ф перестановочным с оператором P:

$$\frac{\partial \mathrm{P}\tilde{\rho}}{\partial t} = \mathrm{P}\Phi\tilde{\rho} = \Phi\mathrm{P}\tilde{\rho}$$

**Приложение G** Невозможность **самопредсказания системы.**
Предположим, что существует мощная вычислительная машина способная предсказать будущее свое и окружения на основе расчета движения всех молекул.Пусть ее предсказания - выкатывание черного или белого шара из некого устройства, входящего в одну систему с машиной и описываемого ею. Устройство выкатывает белый шар, если машина предсказывает черный и наоборот черный шар при предсказании белого. Ясно, что предсказания машины всегда не верны. Так как выбор окружающей среды произволен, следовательно этот контрпример доказывает невозможность точного самонаблюдения и самовычисления. Так как результат действия устройства всегда противоречит предсказаниям компьютера, следовательно полное самопредсказание системы, включая и компьютер, и устройство невозможно.

**Приложение H Соответствие квантовой и классической механик. Таблица.**

| **Квантовая механика** | **Классическая механика** |
|---|---|
| Матрица плотности | Функция фазовой плотности |
| Уравнение движения для матрицы плотности | Уравнение Лиувилля |
| Редукция волнового пакета | Огрубление функции фазовой плотности или гипотеза молекулярного хаоса |
| Неустранимое воздействие наблюдателя или окружения, описываемое редукцией | Теоретически бесконечно малое, но в реальности конечное взаимодействие системы с наблюдателем или окружением |
| Ненулевые недиагональные элементы матрицы плотности | Корреляции между скоростями и положениями молекул в разных частях системы |